\documentclass{aa}  

\usepackage{graphicx}
\usepackage{txfonts}
\usepackage{lipsum}
\usepackage{subcaption}         
\usepackage{lscape}             
\usepackage{placeins}           

\usepackage{multirow} 
\usepackage{tabularx}
\usepackage{caption}
\usepackage{tikz} 

\usepackage[colorlinks=true,
            linkcolor=blue,
            citecolor=blue,
            urlcolor=blue]{hyperref}
                                
\begin{document}

   \title{Parametrizing the projected wind fields of ultra-hot Jupiters in thermal emission: an application to GCM spectra of WASP-76b}

   \titlerunning{Parametrizing the projected wind fields of ultra-hot Jupiters in thermal emission}



   \author{Joost P. Wardenier\inst{1}\fnmsep\inst{2},
           Romain Allart\inst{2},
           Georgia Mraz\inst{3}\fnmsep\inst{4},
           Louis-Philippe Coulombe\inst{5}\fnmsep\inst{2},
           Frédéric Genest\inst{2}, \\
           Vigneshwaran Krishnamurthy\inst{3}\fnmsep\inst{4},
           Enola Quenet\inst{2},
           \and Vincent Yariv\inst{6}
        }

   \institute{Weltraumforschung und Planetologie, Physikalisches Institut, University of Bern, Gesellschaftsstrasse 6, 3012 Bern, Switzerland\\
   \email{joost.wardenier@unibe.ch}
   \and Institut Trottier de Recherche sur les Exoplanètes, Département de Physique, Université de Montréal, Montréal, Québec, Canada
   \and Trottier Space Institute at McGill, 3550 rue University, Montréal, QC H3A 2A7, Canada
   \and Department of Physics, McGill University, 3600 rue University, Montréal, QC H3A 2T8, Canada
   \and Planétarium de Montréal, Espace pour la Vie, 4801 av. Pierre-de Coubertin, Montréal, Québec, Canada
   \and Institut de Planétologie et Astrophysique de Grenoble, Grenoble, CNRS, IPAG, 38000 Grenoble, France}

   \authorrunning{Joost P. Wardenier et al.}

   \date{Received March 24, 2026 -- Accepted June 21, 2026}


  \abstract
   {High-resolution thermal emission spectroscopy provides a powerful probe of atmospheric circulation in ultra-hot Jupiters (UHJs), with Doppler shifts encoding information about the 3D wind field across the planet disk. Retrieving these wind properties from phase-dependent emission spectra requires a forward model that is both physically motivated and computationally tractable.}
   {We aim to investigate what 3D wind information is contained in the thermal emission spectra of UHJs in the first place, and whether the complicated 3D atmospheric structure of these planets can be captured by a simplified forward model with a small number of parameters.} 
   {{\color{black}{We present \texttt{dopplerkernel}}}, a new forward model that parametrizes the projected line-of-sight velocity field on the planet disk using four wind parameters (an equatorial jet speed $v_\mathrm{jet}$ and width $\sigma_\mathrm{jet}$, {\color{black}{a source-to-sink flow speed}} $v_\mathrm{wind}$, and a flow convergence longitude $\varphi_\mathrm{sink}$) and constructs a broadening kernel via weighted kernel-density estimation. We apply this framework in a Bayesian retrieval to synthetic $K$-band emission spectra of WASP-76b generated from three 3D GCM outputs that span a range of atmospheric drag strengths. We test the model across multiple spectral resolutions and orbital phase coverages.}
   {Our retrievals successfully recover the equatorial jet in the drag-free GCM to within ${\sim}1$~km/s of the zonal mean and infer day-to-night wind speeds in good agreement with GCM averages at 1--10~mbar for all three drag regimes. We find that spectral resolutions of $R \sim 100{,}000$ offer an optimal trade-off: sufficient to resolve global wind features while avoiding spurious detections caused by model--data mismatches at higher resolution. Combining pre- and post-eclipse phases yields more reliable constraints than either alone, particularly on the orientation of the source-to-sink flow. Experiments with more complex weight functions reveal a degeneracy between the velocity field and the thermal weighting, cautioning against overparametrization.}
   {A parametric broadening-kernel model with a small number of physically interpretable parameters can accurately reproduce the phase-dependent line shifts, shapes, and strengths in UHJ emission spectra. However, the fundamental degeneracy between the velocity distribution and the thermal weight function means that adding model complexity does not necessarily improve wind inference and may introduce spurious features. Bayesian model comparison is essential when applying this framework to real observations.}


   \maketitle
   \nolinenumbers

\section{Introduction}

Ground-based high-resolution spectroscopy (\citealt{Birkby2018,Snellen2025}) provides a direct window into the atmospheric circulation of hot gas-giant exoplanets, most notably ultra-hot Jupiters (UHJs; \citealt{Arcangeli2018,Parmentier2018}). Measurements of a planet's wind profile can be made in transmission (\citealt{Snellen2010,Louden2015,Flowers2019a,Gandhi2022,Seidel2025}) or thermal emission (\citealt{Pino2022,costasilva2024,Lesjak2024,Guilluy2025,Bazinet2025}), based on the observed Doppler shifts of spectral lines. While transmission spectroscopy probes the terminator region of the atmosphere, emission spectroscopy captures the flux emitted by the entire planet disk. This flux tends to emerge from deeper atmospheric layers compared to those probed during a transit. Therefore, both types of observations yield complementary insights into an exoplanet’s circulation pattern.

Wind-speed measurements on UHJs are relevant for multiple reasons. Due to their tidally-locked nature, UHJs exhibit extreme day-night temperature contrasts, and wind speeds determine how efficiently heat is redistributed from the hot dayside to the cooler nightside. Atmospheric circulation also governs temperature gradients in the terminator region, especially via H$_2$ dissociation/recombination and the resulting latent-heat release (\citealt{Bell2018,Tan2019,Roth2021,Tan2024,Wardenier2024}). Furthermore, winds dictate the chemical transport of species from the dayside to the nightside, sculpting the chemical composition of the terminator (\citealt{Agundez2012,Mendonca2018,Baeyens2024,Tsai2024}). Finally, the wind profiles of UHJs are impacted by the planetary magnetic field, as temperatures are high enough to produce ionized flows of material (\citealt{Beltz2022,Beltz2023,Beltz2024,Savel2024, Seidel2026}). Hence, wind-speed measurements enable constraints on the magnetic-field strengths of UHJs. All in all, obtaining a more detailed picture of the atmospheric circulation of gas giants is key to understanding their climate, composition, and energy balance. 

So far, the most robust wind-speed measurements on gas giants have been made from transmission observations. This is because winds flow essentially parallel to the line of sight in transit geometry, resulting in the strongest possible Doppler shifts (typically on the order of 1--10 km/s). At high resolution, it is common practice for atmospheric retrievals to include free parameters (e.g., $\Delta K_\text{p}$, $\Delta V_\text{sys}$) that account for these effects. Moreover, various studies explicitly parametrized contributions from planet rotation, the equatorial jet, and the day-to-night flow at the terminator to build ``broadening kernels'' that describe the absorption-line shapes (\citealt{Brogi2016,Gandhi2022,Gandhi2023,Boucher2023,Maguire2024,Nortmann2025}). While all of these works relied on cross correlation with a model template (\citealt{Snellen2010,Brogi2019}), it is also possible to retrieve wind properties from individually resolved lines (\citealt{Allart2018,Seidel2019a,Seidel2021}), which probe lower atmospheric pressures and, therefore, a different circulation regime. Another approach is to compare high-resolution observations to outputs of 3D global circulation models (GCMs) to broadly constrain wind patterns and atmospheric drag (\citealt{Flowers2019a,Beltz2020,Wardenier2021,Wardenier2024,Savel2022}).

In thermal emission, the signature of atmospheric winds is more subtle as the observed Doppler shifts are an ``average'' over the entire planet disk. Also, because winds on UHJs are zonal/meridional at the pressures probed in thermal emission, their line-of-sight components converge to zero towards the center of the disk. In \citet{Wardenier2025} we showed that planetary rotation is the most important driver of Doppler shifts in the emission spectra of UHJs, producing a lower apparent orbital velocity, $K_\text{p}$ (see also \citealt{Hoeijmakers2022,Sing2024,Bazinet2025,Guilluy2025}). When it comes to retrieving wind profiles in emission, \citet{Lesjak2024} presented an intuitive approach. Their retrieval relied on sampling a parametrized velocity field on a 2D grid and building a histogram of line-of-sight velocities to obtain the corresponding broadening kernel -- akin to work by \citet{Brogi2016} for transmission. Using this framework, they inferred a global day-to-night flow of $\sim$4 km/s from post-eclipse observations of WASP-189b with VLT/CRIRES+, while finding no statistical evidence for an equatorial jet. More recent work by \citet{Zhang2026} revealed evidence for strong winds on KELT-9b based on post-transit and pre-eclipse observations from Keck/KPF.

While wind measurements from UHJ emission spectra are tantalizing, it is important to state a few caveats. Wind measurements are based on \emph{residual} Doppler shifts in the planet rest frame, after correcting for the systemic velocity of the star and the orbital motion of the planet. This means that imprecise knowledge of the planet's ephemeris (i.e., period, mid-transit time, eccentricity) can bias these Doppler shifts and lead to spurious ``detections'' of atmospheric circulation -- see \citet{Smith2023} for a detailed discussion. In addition, the uncertainty in the orbital velocity of UHJs can be multiple km/s (\citealt{Wardenier2025,Bazinet2025}), hampering the interpretation of anomalous signals. One challenge in this regard is that many UHJs orbit rapidly rotating A-type stars with different levels of activity, making it hard to obtain precise RV semi-amplitudes. Also, host-star masses reported in the literature can vary (see the case of WASP-121\footnote{Stellar masses of WASP-121 reported in the literature include {1.358 $\pm$ 0.08 $M_\odot$} (\citealt{Delrez2016}), {1.38 $\pm$ 0.02 $M_\odot$} (\citealt{Borsa2021}), {1.33 $\pm$ 0.019 $M_\odot$} (\citealt{Sing2024}), and {1.42 $\pm$ 0.03 $M_\odot$} (\citealt{Prinoth2025}). The mass from \citet{Sing2024} may be slightly biased as it was derived from the apparent $K_\text{p}$ of WASP-121b, which also includes a contribution from planet rotation (see \citealt{Wardenier2025}).}), leading to different expected $K_\text{p}$ values for the same planet. Finally, obtaining an unbiased measurement of $V_\text{sys}$ is not always trivial due to different approaches in data processing -- see the analyses of the same dataset by \citet{costasilva2024} and \citet{Guilluy2025} -- and the fact that instrument velocity offsets may change between different visits (see also \citealt{Lenhart2026}). In summary, even though atmospheric circulation \emph{should} manifest itself in the emission spectra of UHJs, it is not always obvious to what extent residual Doppler shifts can be reliably attributed to winds. This underscores the need for broadening-kernel models that are more physically motivated, and that model line shifts and line shapes in a more self-consistent way.

\begin{figure*}
        \centering
        \makebox[\textwidth][c]{\hspace{-50pt} \includegraphics[width=1.22\textwidth]{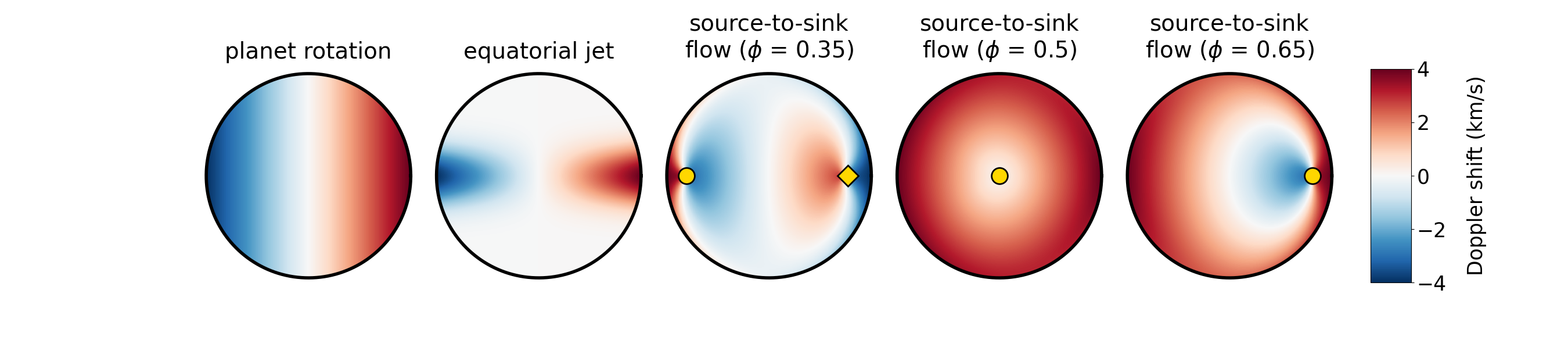}}
        \vspace{-35pt}
        \caption{Components of the velocity-field parametrization discussed in Section \ref{sec:parametric_model}. Each panel shows the line-of-sight velocities $v_{\text{LOS}}$ on the planet disk, where $v_{\text{LOS}} < 0$ km/s implies a blueshift and $v_{\text{LOS}} > 0$ km/s implies a redshift. The contributions from planet rotation and the equatorial jet (two left panels) are independent of orbital phase, while the projection of the source-to-sink flow changes over the course of the orbit (three right panels). {\color{black}{The source at the subtellar point is indicated by the yellow circles, while the sink is marked by the yellow diamond.}} At $\phi = 0.5$, the source-to-sink flow only induces a redshift, but at other orbital phases some regions on the disk become blueshifted. For this plot, we used $v_\text{rot} = v_\text{jet} = v_\text{wind} =$ 4 km/s, $\sigma_\text{jet} = 0.2 \times R_\text{disk}$, and $\varphi_\text{sink} = 100^\circ$. Note that the projected source-to-sink flows at $\phi = 0.35$ and $\phi = 0.65$ would have been each other's mirror images for $\varphi_\text{sink} = 180^\circ$. However, in this example, the flow converges to a sink that is offset from the antistellar point, producing an asymmetry.}
        \label{fig:doppler_components}
\end{figure*}

\begin{figure*}[t]
        \centering
        \vspace{-40pt}
        \makebox[\textwidth][c]{\hspace{-20pt} \includegraphics[width=1.17\textwidth]{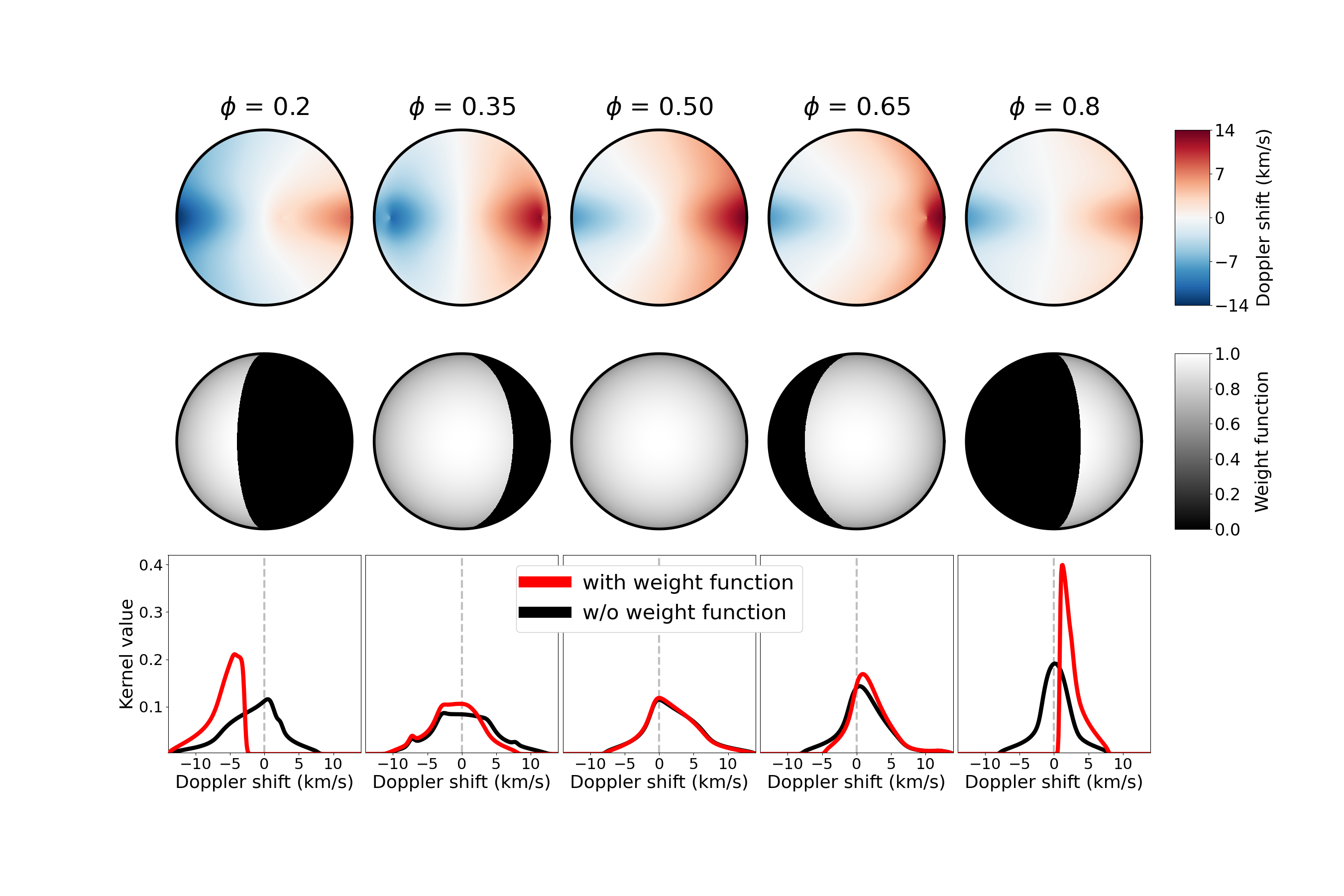}}
        \vspace{-60pt}
        \caption{Example of a velocity-field parametrization and the resulting broadening kernels. \textbf{Top row:} 2D projection of the velocity field at different orbital phases $\phi$, with \mbox{$v_\text{rot}$ = 5 km/s}, $v_\text{jet}$ = 6 km/s, $v_\text{wind}$ = 3 km/s, $\sigma_\text{jet} = 0.2 \times R_\text{disk}$, and $\varphi_\text{sink} = 120^\circ$. \textbf{Middle row:} 2D weight function that is zero on the nightside and nonzero on the dayside. {\color{black}{On the dayside, we assume a linear limb-weighting profile with \mbox{$w_0$ = 2} (such that the center is weighted twice as heavily as the edge of the planet disk).}} \textbf{Bottom row:} Broadening kernels computed with SciPy's kernel-density estimation function (\texttt{scipy.stats.gaussian\_kde}). The red kernels were generated using the weight function from the middle row, while the black kernels assume equal weights across the entire planet disk. The smaller the dayside region that is in view, the larger the discrepancy between both broadening kernels. All kernels are normalized such that the integrated area is unity. {\color{black}{In this way, flux is conserved when convolving the kernel with a given spectrum.}}}
        \label{fig:doppler_kernel_figure}
\end{figure*}

In this work, we will take a step back and investigate what ``3D wind information'' is contained in the emission spectra of UHJs in the first place, and whether it is possible to capture their complicated 3D atmospheric structure in a simplified forward model with a small number of parameters. To this end, we use a parametric retrieval framework inspired by \citet{Lesjak2024} and apply it to synthetic phase-dependent emission spectra of WASP-76b, generated from 3D GCM outputs. In Section \ref{sec:methods}, we discuss the parametric broadening-kernel model, the GCM outputs, and the retrieval setup. In Section \ref{sec:results}, we present the results of our retrieval experiments and discuss their implications. Finally, Section \ref{sec:conclusion} provides a summary and conclusion. 

\section{Methods}
\label{sec:methods}

\subsection{Broadening-kernel model}
\label{sec:parametric_model}

In this section, we describe \texttt{dopplerkernel}\footnote{The Python code and a Jupyter Notebook tutorial are available on \href{https://github.com/joostwardenier/dopplerkernel}{\texttt{github.com/joostwardenier/dopplerkernel}}}, our forward model to generate broadening kernels that capture the impact of planet rotation and atmospheric dynamics on the thermal emission spectra of UHJs. We take the model from \citet{Lesjak2024} as a starting point (see their Fig. 6). The idea is to parametrize the projected line-of-sight velocities on the planet disk, sample them on a grid\footnote{In their work, \citet{Lesjak2024} use a grid of 2001 $\times$ 2001 cells.}, assign a weight to each grid cell that mimics the effect of the temperature structure, and build a histogram of line-of-sight-velocity values. This histogram is equivalent to the broadening kernel associated with the 2D velocity field. 

To limit the number of parameters in the model, we restrict ourselves to three easy-to-interpret contributions, which are shown in \mbox{Fig. \ref{fig:doppler_components}}. First, we account for solid-body rotation, where $v_\text{rot}$ is the rotational velocity of the planet at the equator. Second, we include an equatorial jet with a Gaussian velocity profile as a function of latitude. $v_\text{jet}$ is the jet speed at zero latitude, while $\sigma_\text{jet}$ (the standard deviation of the Gaussian) controls how fast the jet speed drops off away from the equator. {\color{black}{The third component is a global ``source-to-sink'' flow with velocity $v_\text{wind}$ that originates from the substellar point and converges to a different point on the equator with longitude $\varphi_\text{sink}$.}} After much trial and error, we found that including $\varphi_\text{sink}$ is necessary to reproduce the magnitude of the blueshifts seen in pre-eclipse. The source-to-sink flow becomes a purely substellar-to-antistellar flow when $\varphi_\text{sink} = \pm 180^\circ$. In Appendix \ref{app:A_0}, we present the equations describing our parametrization of the velocity field.

While the 2D projections for planet rotation and the equatorial jet are independent of orbital phase, the 2D projection of the source-to-sink flow changes as the tidally locked planet orbits its host star. \mbox{Fig. \ref{fig:doppler_components}} depicts three such projections at orbital phases $\phi = 0.35$ (pre-eclipse), $\phi = 0.5$ (secondary eclipse), and $\phi = 0.65$ (post-eclipse), respectively, assuming $\varphi_\text{sink} = 100^\circ$. In this example, the source-to-sink flow converges to a point on the nightside just beyond the terminator plane.

Because UHJs are tidally locked, it is also important to account for the day-night contrast in the model. While the thermal inversion on the dayside produces strong emission features, the vertical temperature profile on the nightside is non-inverted and more isothermal, giving rise to much shallower absorption features. In \citet{Wardenier2025}, we showed that the nightside absorption signal of UHJs can be an order of magnitude smaller than their dayside emission signal, depending on the atmospheric drag strength. Therefore, the simplest approximation (especially in pre-and post-eclipse phases) is to use a weight function that is zero on the nightside. On the dayside, we apply a linear weight profile with a value of $w_0$ at the disk center and a value of 1 at the disk edge. {\color{black}{In this parametrization, $w_0 > 1$ means that the disk is more \emph{center-weighted}, while $0 < w_0 < 1$ implies that the disk is more \emph{limb-weighted}. We explicitly do not use the terms ``limb darkening'' and ``limb brightening'' here because these are normally associated with changes in absolute flux across the disk. In high-resolution emission spectroscopy, however, the weight function should encode changes in line strength. That is, the local flux \emph{difference} between a line core and the spectral continuum.}}

Fig. \ref{fig:doppler_kernel_figure} shows what the 2D velocity field and the 2D weight function look like for a given set of parameter values (see caption) at five different orbital phases. We compute the corresponding broadening kernel using the kernel-density estimation function (\texttt{scipy.stats.gaussian\_kde}) from SciPy (\citealt{Virtanen2020}), which takes both the velocity field and the weight function as arguments. The advantage of this function is that the shape of the kernel is robust under different grid sizes, so there is no need to sample the velocity field at very high spatial resolution, which speeds up the calculation. In Appendix \ref{app:B}, we show the impact of different grid sizes on the precision of the broadening kernel and the runtime of the forward model. 

Finally, we normalize the kernel so that its integrated surface area is unity. This step ensures that the total flux is conserved when performing convolution with a given spectrum. {\color{black}{For UHJs, we recognize that the total flux and the line strength also change with orbital phase due to the day-night contrast, so one could opt to rescale the broadening kernel to account for this (see Section \ref{sec:retrieval_setup}). However, we envision that \texttt{dopplerkernel} will mainly be used as an extension to existing retrieval frameworks that may already include a phase-dependent scale factor. Since convolution is a linear operation, scaling the spectrum before or after convolution yields the same result as scaling the kernel. Therefore, it is always possible to apply this step separately if necessary.}}

\begin{figure*}[t]
        \centering
        \vspace{-1pt}
        \makebox[\textwidth][c]{\includegraphics[width=1.01\textwidth]{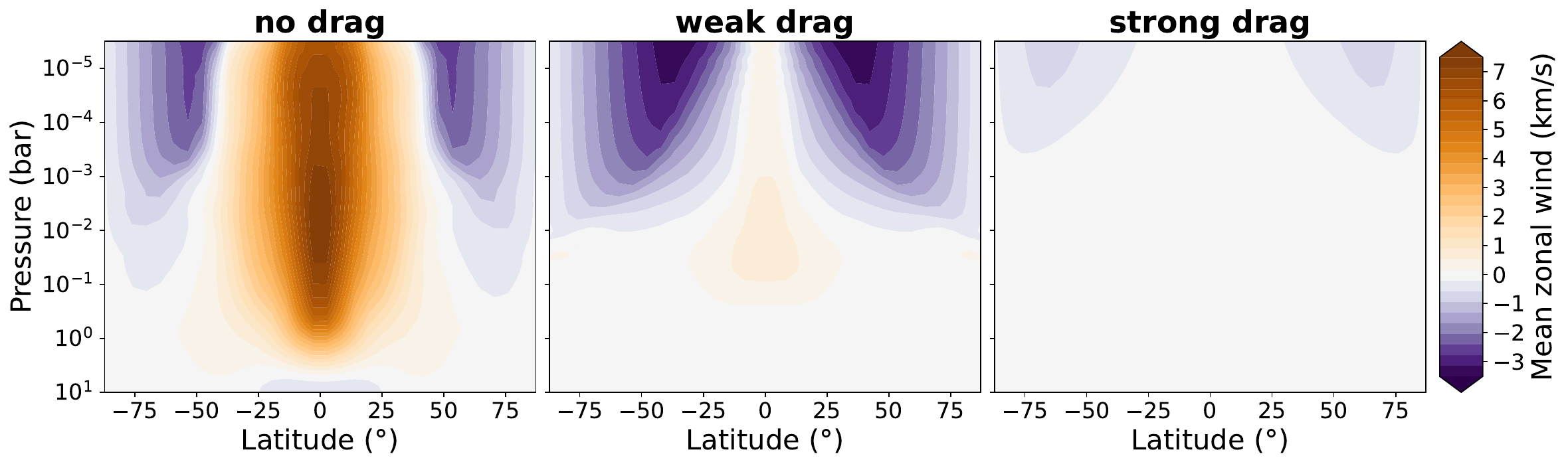}}
        \vspace{-17pt}
        \caption{Mean zonal winds for the three WASP-76b models, obtained by averaging the west-to-east (zonal) wind component over all longitudes. The drag-free GCM (left panel) features a strong equatorial jet, while the others are dominated by a day-to-night flow.}
        \label{fig:zonal_winds}
\end{figure*}

\subsection{Calculating GCM spectra and templates}

In this work, we consider three 3D GCM outputs of the canonical UHJ WASP-76b that were previously used in \citet{Wardenier2021,Wardenier2023,Wardenier2025}. One of the models is drag-free, while the other two have uniform drag timescales $\tau_\text{drag}$ = \mbox{$10^4$ s} (strong drag) and $\tau_\text{drag}$ = \mbox{$10^5$ s} (weak drag), respectively. The drag timescale represents various physical processes in the atmosphere, such as turbulent mixing (\citealt{Li2010}), Lorentz-force braking of ionized winds in the planet's magnetic field (\citealt{Perna2010}), and Ohmic dissipation (\citealt{Perna2010a}). The stronger the atmospheric drag, the less time it takes for an air parcel to lose its kinetic energy, resulting in lower wind speeds and less efficient heat redistribution between the planet's dayside and nightside.

All models were calculated using the SPARC/MITgcm (\citealt{Showman2009,Parmentier2018}), which was recently renamed ADAM (ADvanced Atmospheric MITgcm; see \citealt{Mehta2025}). Fig. \ref{fig:zonal_winds} shows the mean zonal winds associated with the three GCM outputs. Notably, the drag-free model features a strong equatorial jet, while the models with drag are dominated by day-to-night winds. We refer to \citet{Wardenier2021,Wardenier2023,Wardenier2025} for more detailed descriptions and visualizations of the three WASP-76b models (temperature structures, abundance profiles, line-of-sight-velocity maps, etc.).

We follow the methods discussed in \citet{Wardenier2025} to generate phase-dependent emission spectra between orbital phases $\phi = 0.25$ and $\phi = 0.75$, with  $\Delta\phi = 1/36$ intervals (corresponding to 10 degrees of planet rotation). We use the 3D Monte-Carlo radiative transfer code gCMCRT (\citealt{Lee2022}) to simulate the thermal flux emerging from the planet. gCMCRT accounts for the 3D thermochemical structure of the atmosphere, as well as the Doppler shifts imparted on the opacities by planet rotation and the 3D wind profile. Radiative transfer is performed in the $K$-band between 2.28 and 2.49 $\mu$m, at a spectral resolution $R$ = 500,000. We select this wavelength range as it covers a large number of strong CO lines that are highly suitable for wind measurements (\citealt{Yan2022_CO,vanSluijs2022,Smith2024,Lesjak2024,Pelletier2025,Ramkumar2025}), similar to Fe in the optical. In the radiative transfer, we include opacities from CO (\citealt{Li2015}) and H$_2$O (\citealt{Polyansky2018}), as well as continuum opacities due to collision-induced absorption (CIA) by H$_2$-H$_2$ and H$_2$-He, and bound-free and free-free transitions of H$^-$ (for references, see Table 2 in \citealt{Lee2022b}).

Finally, we generate three ``static'' 1D template spectra based on the average vertical structure of each model's dayside (assuming zero line-of-sight velocity). Such phase-independent templates are similar to spectra obtained from 1D radiative-transfer codes such as CHIMERA (\citealt{Line2013}) or petitRADTRANS (\citealt{Molliere2019a}). {\color{black}{We note that assuming a \emph{non-weighted} average of the 3D GCM dayside may not necessarily be the optimal choice (e.g., \citealt{Blecic2017,Wiser2026}). However, since different dayside regions are in view at different orbital phases, no single 1D template can serve as the best-possible representation of the planet spectrum at all phases simultaneously.}}

\begin{table}[t]
\caption{{\color{black}{Parameters used in the broadening-kernel retrievals and their prior distributions.}}}
\vspace{0pt}
\label{tab:priors}
\centering
\small
\renewcommand{\arraystretch}{1.3}
\begin{tabular}{lp{4.5cm}l}
\hline\hline
\textbf{Parameter} & \textbf{Description} & \textbf{Prior} \\
\hline
$v_{\mathrm{wind}}$ [km/s]      & Source-to-sink-flow speed          & $\mathcal{U}(0,10)$ \\
$\varphi_{\mathrm{sink}}$ [deg] & Sink longitude                     & $\mathcal{U}(-180,180)$ \\
$v_{\mathrm{jet}}$ [km/s]       & Jet speed at equator                          & $\mathcal{U}(0,10)$ \\
$\sigma_{\mathrm{jet}}$ [$R_p$] & Jet spatial width                  & $\mathcal{U}(0,0.5)$ \\
$w_0$                           & Disk-center weight relative to limb & $\mathcal{U}(0,5)$ \\
$\alpha_0$                      & Scale-factor amplitude             & $\mathcal{U}(0,2)$ \\
$\phi_0$                        & Scale-factor offset                & $\mathcal{U}(-0.2,0.2)$ \\
$\log(\beta)$                   & Jitter term                        & $\mathcal{U}(-5,2)$ \\
\hline
\end{tabular}
\tablefoot{$\mathcal{U}(a,b)$ denotes a uniform distribution with lower bound $a$ and upper bound $b$.}
\end{table}

\subsection{Retrieval setup}
\label{sec:retrieval_setup}

To find the best set of wind-field parameters for each of the three GCM outputs, we seek to maximize the log-likelihood between the GCM spectra at $n_\phi$ different orbital phases and the Doppler-broadened 1D template. This gives

\begin{equation}
    \ln \mathcal{L}(\vec{\theta}) = -\frac{1}{2} \sum_{i = 1}^{n_\phi} \sum_{j = 1}^{n_\lambda} \left[ \frac{\big(y_{i,j} - m_j(\vec{\theta}, \phi_i)\big)^2}{\beta^2} + \ln(2 \pi \beta^2) \right],
\end{equation}

\noindent where $n_\lambda$ is the number of wavelength points, $y$ is the GCM flux, $\vec{m}(\vec{\theta},\phi_i)$ is the Doppler-broadened model at phase $\phi_i$ for a set of free parameters $\vec{\theta}$, and $\beta$ is a jitter term. We include the jitter term to absorb small mismatches between the (simplified) model and the more complicated GCM spectra, {\color{black}{resulting in a smoother likelihood function}}.

\begin{figure*}
        \centering
        \vspace{-1pt}
        \makebox[\textwidth][c]{\hspace{-25pt}\includegraphics[width=0.9\textwidth]{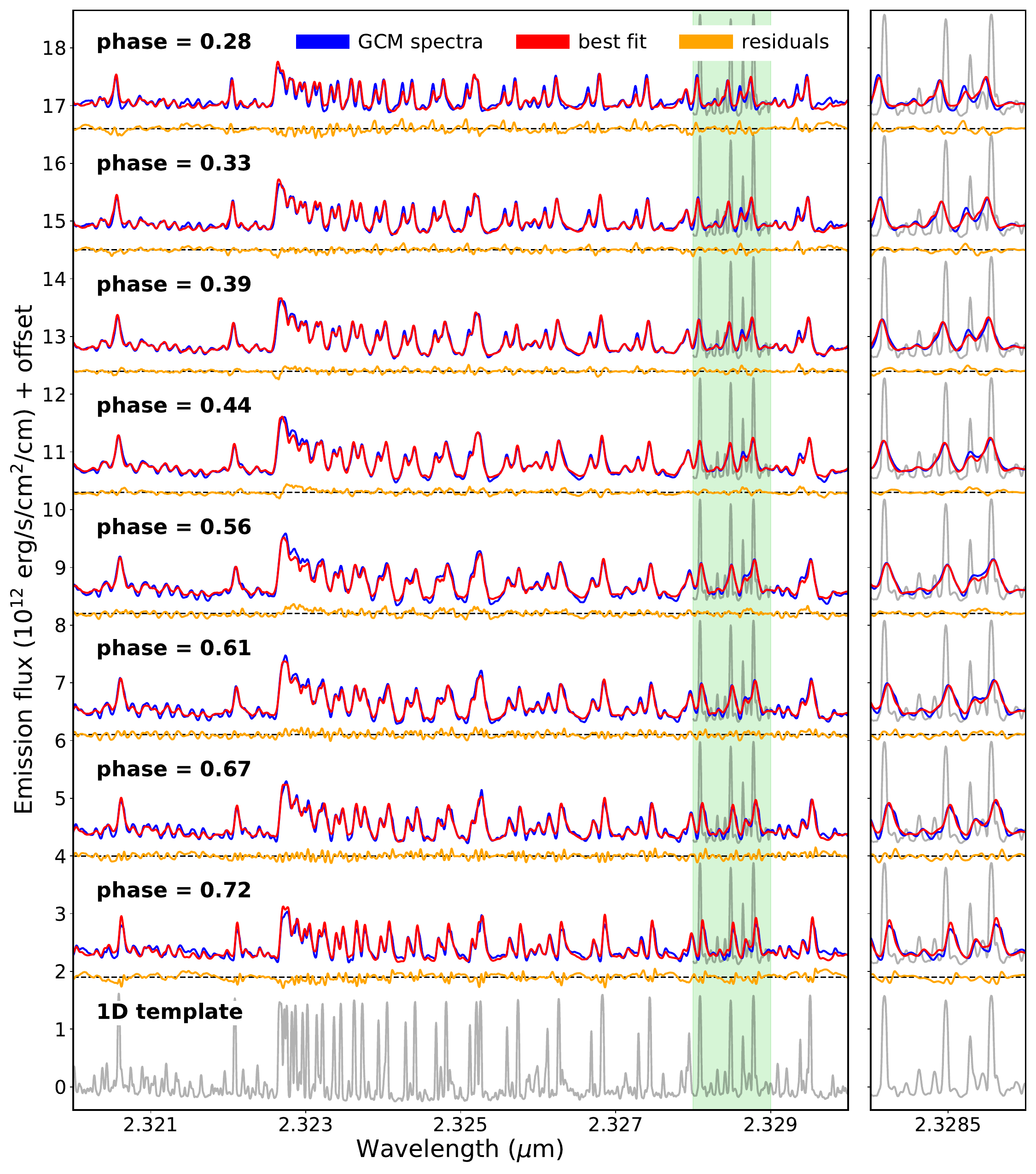}}
        \vspace{-17pt}
        \caption{Comparison of the drag-free GCM spectra (in blue) to the spectra associated with the best-fit broadening kernels (in red) obtained from the $R$ = 500,000 retrieval covering both pre-eclipse and post-eclipse phases. Residuals (in yellow) show the difference between the model and the GCM spectra at each wavelength. {\color{black}{In the bottom of the figure, we show the 1D template spectrum (in grey) that is convolved with a broadening kernel at every orbital phase to produce the model spectra (see Section \ref{sec:retrieval_setup}).}} The vertical green band shows the wavelength range that is plotted in the narrow panel on the right. This panel shows how the Doppler shifts of the GCM spectra and the best-fit spectra change with orbital phase, compared to the static 1D template spectrum in the background.}
        \label{fig:best_fit_spectra}
\end{figure*}

Before starting the retrieval, we place the GCM spectra and the 1D templates on the same ``flat'' continuum level by convolving them with a broad Gaussian and subtracting the result from the original spectra {\color{black}{(see Fig. \ref{fig:best_fit_spectra})}}. In addition, we need to account for the fact that the line strengths in the GCM spectra change as a function of phase -- not just because of changes in Doppler broadening, but mainly because of the rotating 3D temperature structure (see Figs. \mbox{5-8} in \citealt{Wardenier2025}). {\color{black}{Therefore, we multiply the Doppler-broadened templates by a phase-dependent scale factor $\alpha$, which we parametrize as

\begin{equation}
\label{eq:scalefactor}
    \alpha(\phi_i) = \alpha_0 \sin^2\Big(\pi(\phi_i + \phi_0)\Big), 
\end{equation}

\noindent where $\alpha_0$ is the amplitude and $\phi_0$ is a phase offset. This equation is similar to parametrizations used in previous high-resolution studies (\citealt{Herman2022,Ridden-Harper2023,Hoeijmakers2022,Zhang2026}), assuming a day-night contrast ratio of unity. As mentioned previously, we could have incorporated the scaling into the broadening kernel itself, but we prefer to keep things separated for clarity.}}

Our model now has the following form:

\begin{equation}
\label{eq:model}
    \vec{m}(\vec{\theta}, \phi_i) = \alpha(\vec{\theta}, \phi_i) \times \Big[ \vec{t}_\text{1D} \star \vec{k}(\vec{\theta}, \phi_i) \Big], 
\end{equation}

\noindent where the square brackets contain the convolution of the 1D template $\vec{t}_\text{1D}$ with the broadening kernel $\vec{k}$, and $\alpha$ represents the scale factor. {\color{black}{Using \texttt{dopplerkernel}, we sample the planet disk on a 100$\times$100 grid.}} In total, we use eight free parameters in our retrieval: five parameters describing the broadening kernel ($v_\text{jet}$, $\sigma_\text{jet}$, $v_\text{wind}$, $\varphi_\text{sink}$, and $w_0$), two parameters related to the scale factor ($\alpha_0$ and $\phi_0$), and a jitter term $\beta$. Since the equatorial rotation velocity of our models is known ($\sim$5.5 km/s at the photosphere), we choose to keep it fixed. Also, since the main goal of this work is to test our model's ability to capture spectral features of a 3D atmosphere, we do not consider any noise sources. {\color{black}{Future work could involve injection tests with real datasets to test how well certain parameters can be recovered under different conditions}}. 

To sample parameter space and obtain posterior distributions, we use the \texttt{emcee} package (\citealt{Foreman-Mackey2012}). Sampling is run for 300 steps with 100 walkers following a burn-in phase of 100 steps. For each parameter, we assume uniform priors, which are listed in Tables \ref{tab:priors}. We perform retrievals for three different spectral resolutions (\mbox{$R$ = 500,000}, \mbox{$R$ = 100,000}, and \mbox{$R$ = 50,000}) and three different phase ranges: pre-eclipse only ($0.25 < \phi < 0.45$), post-eclipse only \mbox{($0.55 < \phi < 0.75$)}, and pre- and post-eclipse combined.

Because our Doppler-broadened model only describes the wind profile on the thermally inverted dayside (that is, the 1D template only contains \emph{emission} features and the nightside weights are zero), it is not designed to fit the lower-flux ``wiggles'' in the spectrum that may contain a mix of absorption and emission features (see also \citealt{Wardenier2025}). {\color{black}{Therefore, when evaluating the likelihood at a given phase, we leave out all wavelengths for which the model flux is lower than $10\%$ of the maximum model flux (we tried thresholds between $10\%$ and $20\%$, which all gave very similar results).}} In Section \ref{sec:accounting_for_nightside}, we discuss how the model could be generalized to also fit nightside absorption features, should these be discernible in an observation.

\section{Results and discussion}
\label{sec:results}

\subsection{Retrieval results: Best-fit spectra, velocity-fields, and broadening kernels}

Fig. \ref{fig:best_fit_spectra} shows a comparison between the GCM spectra from the drag-free model and the best-fit spectra obtained from our retrieval. To convey as much detail as possible, we present the results for the retrieval at \mbox{$R$ = 500,000}. In the plotted wavelength range, the comb of strong emission lines is due to CO. The smaller features in between these lines are due to H$_\text{2}$O. While the strong CO emission lines only probe the thermally inverted dayside of the planet, some of the weaker spectral features may be a combination of absorption and emission\footnote{The drag-free model is the most extreme case when it comes to the mixing of absorption and emission features. The models with drag have more isothermal nightsides, resulting in weaker absorption lines (\citealt{Wardenier2025}).}. Because these were ignored in the retrieval, the likelihood was not optimized to fit the lower-flux parts of the spectra.

\begin{figure*}
        \centering
        \vspace{-14pt}
        \makebox[\textwidth][c]{\hspace{-10pt}\includegraphics[width=1.1\textwidth]{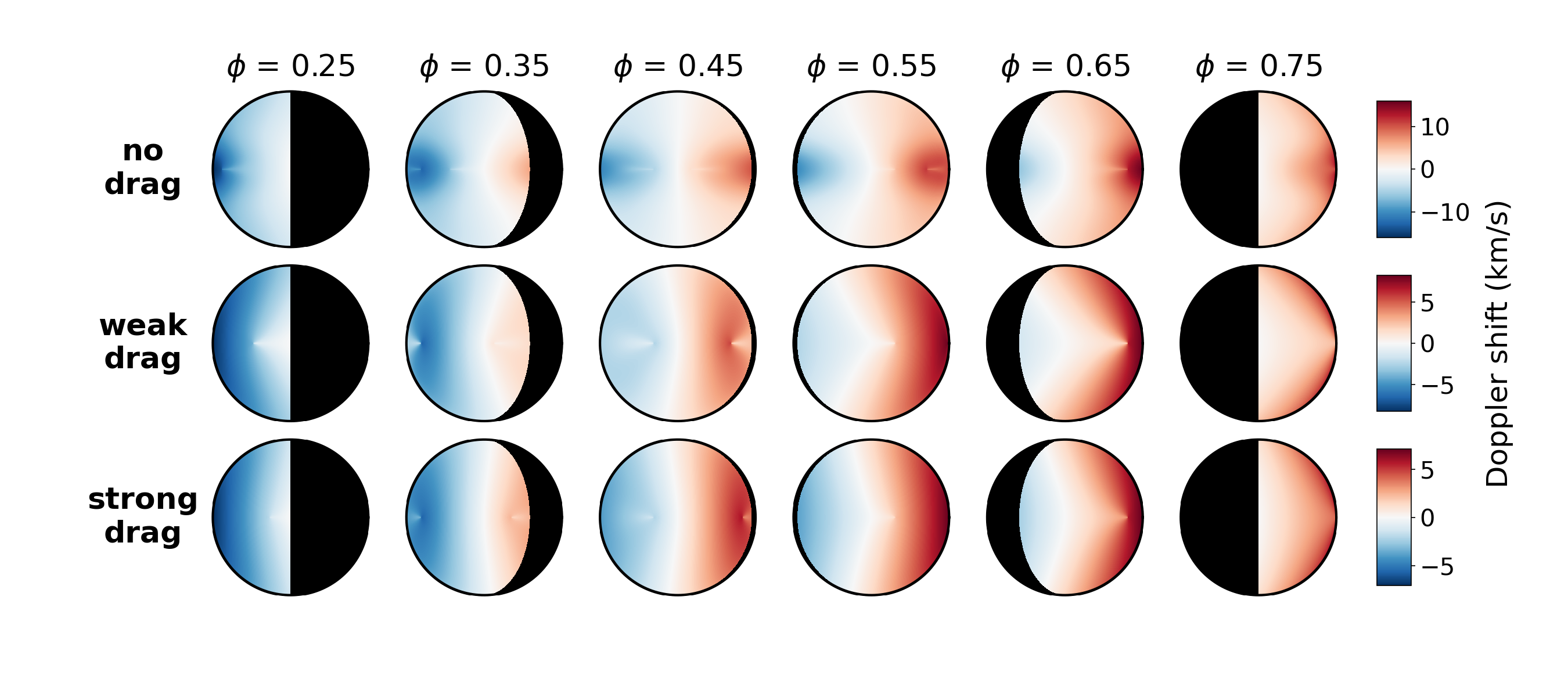}}
        \vspace{-44pt}
        \caption{Best-fit velocity-field projections (winds plus solid-body rotation) for the $R$ = 100,000 retrievals covering both pre-eclipse and post-eclipse phases. Each row depicts the result for a different GCM output of WASP-76b. At each phase $\phi$, the nightside is masked out as our retrievals only recover information from the dayside of the planet. For the models with drag, we set the jet speed to zero as the retrievals converged to negligible $\sigma_{\text{jet}}$ values compared to the drag-free model (see Table \ref{tab:priors_posteriors_resolution}).}
        \label{fig:best_fit_wind_fields}
\end{figure*}

\begin{figure*}
        \centering
        \vspace{-12pt}
        \makebox[\textwidth][c]{\hspace{-10pt}\includegraphics[width=1.1\textwidth]{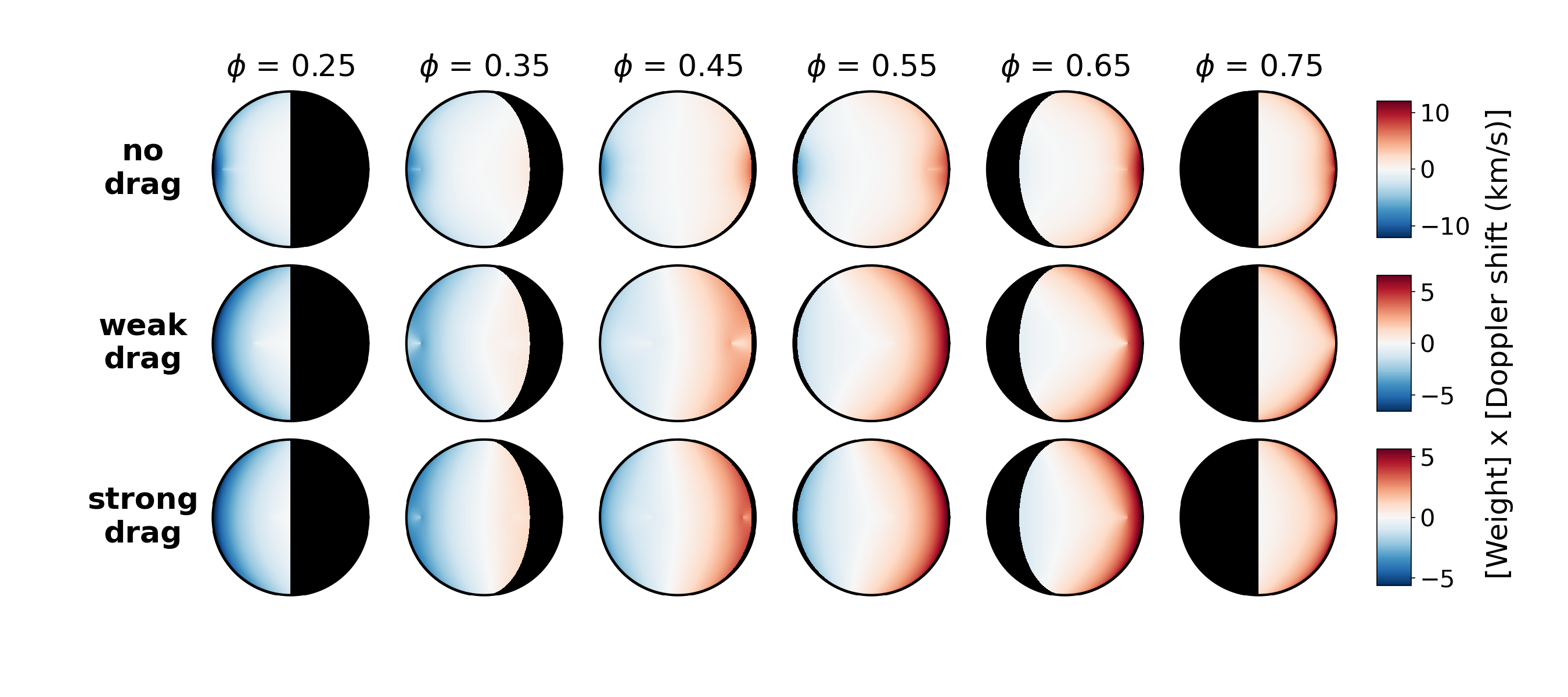}}
        \vspace{-44pt}
        \caption{{\color{black}{Plot similar to Fig. \ref{fig:best_fit_wind_fields}, but now showing the velocity-field projections multiplied by the best-fit weight functions, defined by $w_0$. Note that the meaning of the colormap is different from Fig. \ref{fig:best_fit_wind_fields}, as it now represents a \emph{product} of weights and velocities.}}}
        \label{fig:best_fit_wind_fields_times_weights}
\end{figure*}

As demonstrated in Fig. \ref{fig:best_fit_spectra}, the best-fit model provides an excellent fit to the GCM spectra across all phases, with the residuals being much smaller than the absolute flux variations across the spectra. Towards quadrature, the residuals increase somewhat, as a smaller part of the dayside is in view. However, given that our forward model does not account for the planet's nightside at all, the similarity between the phase-dependent best-fit model and the GCM spectra is still striking. The right panel in Fig. \ref{fig:best_fit_spectra} also shows how the model is able to correctly track the changing Doppler shift across pre-and post eclipse. In pre-eclipse, the model is (blue)shifted to shorter wavelengths compared to the 1D template, while the model is (red)shifted to longer wavelengths in post-eclipse.

\begin{figure*}[t] 
\centering 
\vspace{-2pt} 

\makebox[\textwidth][c]{\includegraphics[width=1.01\textwidth]{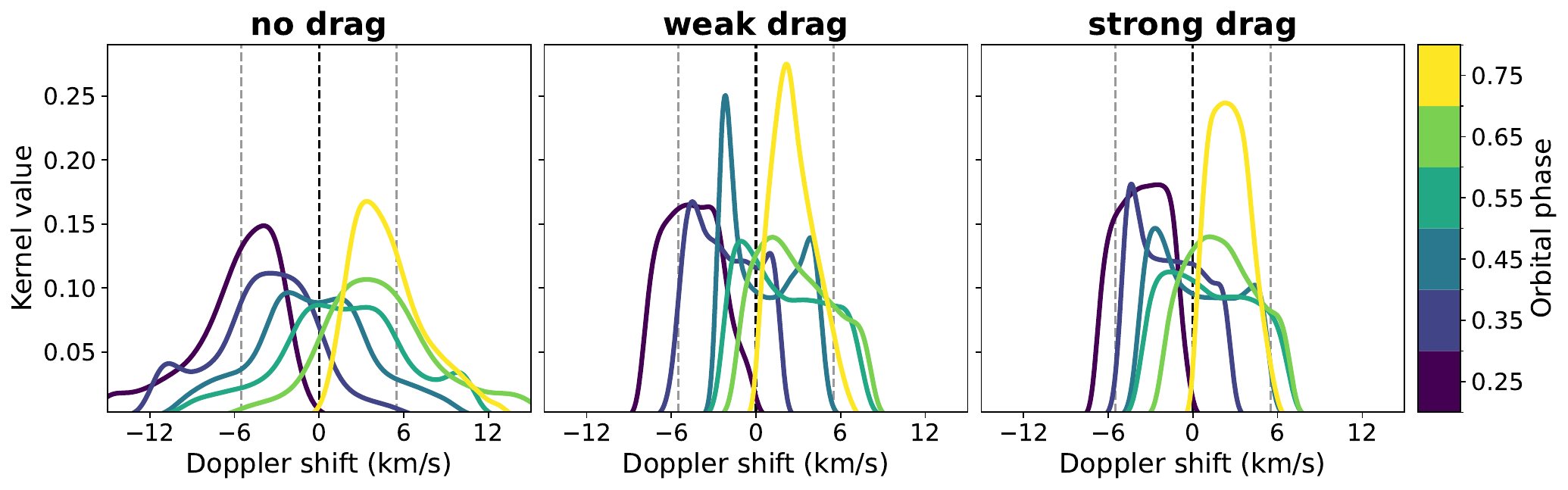}} 
\vspace{-17pt} 

\caption{Best-fit broadening kernels corresponding to the best-fit wind fields shown in Figs. \ref{fig:best_fit_wind_fields} and \ref{fig:best_fit_wind_fields_times_weights}. The grey dashed lines show the equatorial rotation velocity at the photosphere ($\sim$5.5 km/s) in the WASP-76b models. Any Doppler shift in excess of this value cannot be explained by solid-body rotation alone and also requires a contribution from winds.} 

\label{fig:best_fit_kernels} 

\end{figure*}

Fig. \ref{fig:best_fit_wind_fields} depicts the best-fit velocity fields obtained from the retrieval at $R$ = 100,000 (similar to the resolution of a spectrograph like CRIRES+) covering both pre- and post-eclipse. The plots illustrate that planet rotation is still the dominant contributor to the phase-dependent Doppler shifts, as the line-of-sight velocities shift from mostly negative (moving \emph{towards} the observer) at phases 0.25-0.35 to mostly positive (moving \emph{away} from the observer) at phases 0.65-0.75. The winds add extra features on top of this, such as the equatorial jet in the drag-free model. The velocity fields for the models with drag look relatively similar as they only differ in terms of their source-to-sink flow speed ($\sim$1.5 vs. $\sim$3 km/s). The singularities in the maps are the substellar point at zero longitude, from which the winds diverge, and the ``sink'' at longitude $\varphi_\text{sink}$, to which the winds converge. Shifting the sink onto the dayside creates more enhanced blueshifts away from the rotation axis in pre-eclipse, which are needed to fit the Doppler shifts of the GCM spectra. {\color{black}{In Fig. \ref{fig:best_fit_wind_fields_times_weights}, we show similar maps as in Fig. \ref{fig:best_fit_wind_fields}, except that we now multiply the velocity fields by the best-fit weight functions from each retrieval. As we will discuss later in Section \ref{sec:results_posteriors}, all retrievals converge to a more limb-weighted disk ($w_0 \sim 0.05$ for the drag-free model and $\sim$0.3 for the models with drag), where most of the signal comes from the regions near the edge rather than the center.}}  

Fig. \ref{fig:best_fit_kernels} shows the phase-dependent broadening kernels corresponding to the {\color{black}{(weighted) velocity fields from Figs. \ref{fig:best_fit_wind_fields} and \ref{fig:best_fit_wind_fields_times_weights}.}} As explained in Section \ref{sec:parametric_model}, the kernels are histograms of the line-of-sight-velocity values, weighted by the weight function. As expected, the drag-free atmosphere produces the strongest Doppler shifts and a larger amount of broadening compared to the models with drag (indicating a larger velocity dispersion across the planet disk). Overall, it holds that the amount of broadening increases towards the eclipse, when the largest part of the dayside is in view. Another interesting observation is that many of the kernels have non-Gaussian shapes -- e.g., featuring a plateau or a double peak. This demonstrates the ability of our forward model to produce more complicated, phase-dependent kernel shapes that would otherwise be hard to parametrize.

\subsection{Retrieval results: Posterior distributions at R = 100,000}
\label{sec:results_posteriors}

\begin{figure*} 
\centering 
\vspace{-2pt} 
\makebox[\textwidth][c]{\includegraphics[width=0.96\textwidth]{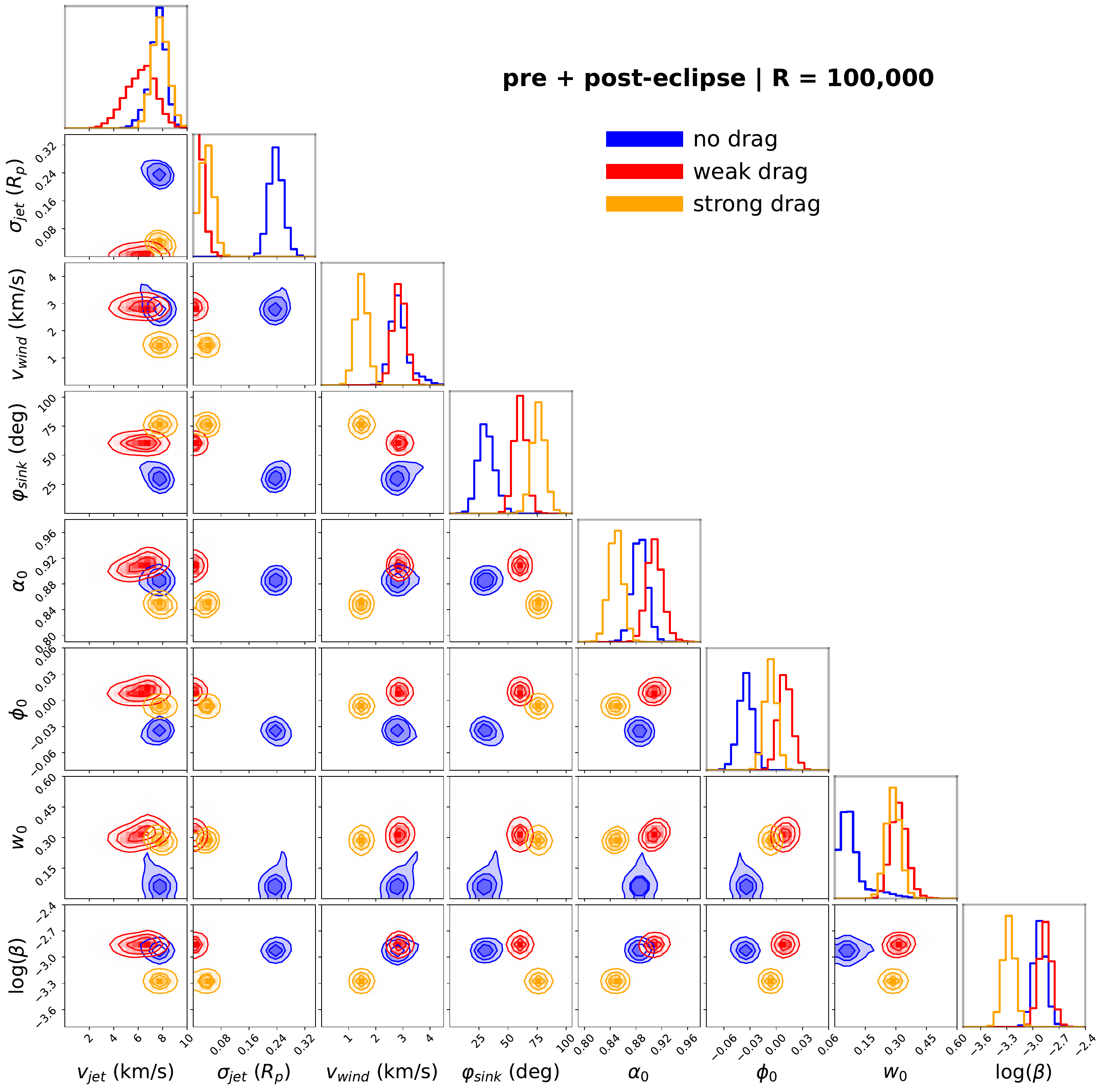}} 
\vspace{-17pt} 
\caption{{\color{black}{Corner plot for the $R$ = 100,000 retrievals covering both pre-eclipse and post-eclipse phases. Each color represents a different GCM output of WASP-76b.}}} 
\label{fig:corner_R1e5_all_phases} 
\end{figure*}

Fig. \ref{fig:corner_R1e5_all_phases} shows the posterior distributions obtained for the retrievals at $R$ = 100,000 that cover both pre- and post-eclipse. {\color{black}{Because our retrievals are performed with ``noiseless data'' (i.e., the spectra generated from the 3D GCM outputs), the widths of the posteriors are less meaningful, as these would normally depend on the signal-to-noise ratio of the observation. However, we still choose to show the posteriors here because they allow for easy visual comparison between the different retrievals and because they can still shed light on potential degeneracies that would otherwise go unnoticed.}} 

For the drag-free model, we recover the equatorial jet at a velocity of \mbox{$\sim$8 km/s}. Conversely, the retrievals for the strong-drag and weak-drag models show no significant evidence of a jet, as the retrieved widths $\sigma_{\mathrm{jet}}$ lie very close to zero. {\color{black}{We should point out, however, that the jet speed still appears to be constrained for these two models, while we would expect it to be \emph{unconstrained} if $\sigma_{\mathrm{jet}}$ was strictly zero. Our hypothesis is that the inference of these narrow, spurious jets is due to two factors: (i) the GCM spectra being noiseless and (ii) our parametric model being a simplified representation of the more intricate 3D wind profile in the GCM. Even when a jet is absent in the GCM, the source-to-sink flow in our broadening-kernel model may not perfectly describe the remaining flow in the GCM (after all, we can only leverage two free parameters, $v_{\mathrm{wind}}$ and $\varphi_{\mathrm{sink}}$). Therefore, the likelihood could increase in a situation where a small, narrow jet can replicate some of the small, subtle features in the spectrum that are left unaccounted for by the source-to-sink flow. In a scenario with noise, these subtle features would be less likely to be discernible\footnote{As we will discuss more elaborately in Section \ref{sec:advanced_weight_function}, any comparison to real data should always include an Occam's razor-like test to verify whether adding more parameters to the model is justified.}.}} When it comes to the source-to-sink flow, the retrieved wind speeds are $\sim$3 km/s for the drag-free and weak-drag models, and $\sim$1.5 km/s for the strong-drag model. {\color{black}{Note that we cannot directly compare these values to the zonal means from Fig. \ref{fig:zonal_winds}, as a purely substellar-to-antistellar flow would produce a zonal mean of zero.}}

\begin{figure}  
\vspace{-8pt} 
{\hspace{-8pt} \includegraphics[width=0.50\textwidth]{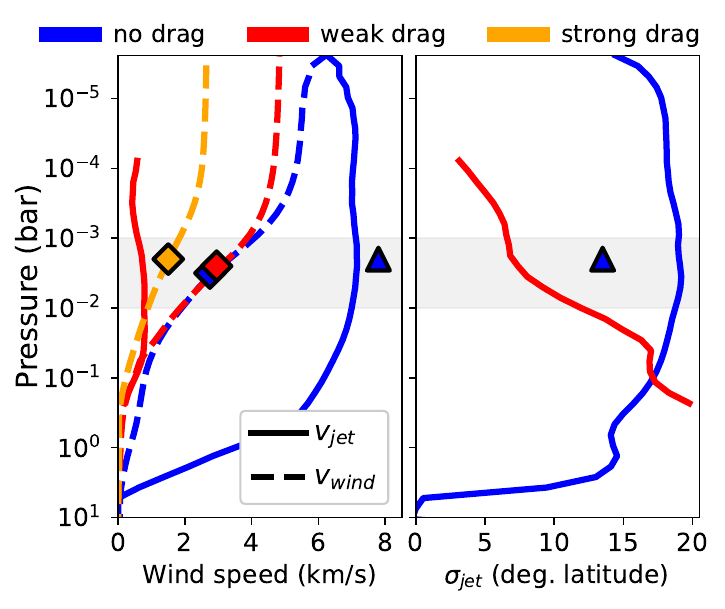}} 
\vspace{-19pt} 
\caption{Comparison of the retrieved wind speeds and jet widths (diamonds and triangles) against pressure-dependent averages from the GCM outputs of WASP-76b (solid and dashed lines). \textbf{Left panel:} Solid lines show the average jet speed of the drag-free and weak-drag models, obtained by fitting Gaussians to the mean zonal winds (see Fig. \ref{fig:zonal_winds}) at every pressure. The dashed lines show the average wind speed on the dayside, obtained by averaging the norms of all wind vectors more than 25$^\circ$ away from the equator. The diamonds show the retrieved day-to-night wind speeds for the three models, plotted at the pressures where they agree best with the GCM averages (in the 1-10 mbar range). The blue triangle indicates the retrieved jet speed for the drag-free model. \textbf{Right panel:} Solid lines show the standard deviation of the Gaussians fitted to the zonal winds. The blue triangle indicates the retrieved value of $\sigma_\text{jet}$ for the drag-free model.} 
\label{fig:1D_velocity_profiles} 
\end{figure}

To make a more precise comparison between the GCM outputs and the retrievals, Fig. \ref{fig:1D_velocity_profiles} shows the average jet properties and wind speeds of the three GCM outputs with the retrieved values superimposed. {\color{black}{To compute the jet speeds and widths for each GCM, we fit Gaussians to the mean zonal winds from Fig. \ref{fig:zonal_winds} at every pressure. To obtain a value representing the speed of the source-to-sink flow, we average the norms of all wind vectors on the GCM daysides that are more than 25$^\circ$ away from the equator (excluding the regions where the jet may persist). While there is no guarantee that this number is the best representation of $v_\text{wind}$, we choose this approach because it is simple. Also, at latitudes beyond $\pm$$25^\circ$, the source-to-sink flow is essentially the only wind component in our broadening-kernel model, so $v_\text{wind}$ is the only parameter that can capture the GCM wind speeds in this region.}}

The left panel in Fig. \ref{fig:1D_velocity_profiles} demonstrates that the retrieved source-to-sink flow speeds agree very well with the average GCM wind speeds between 1 and 10 mbar, which are typical pressures probed in thermal emission. In the drag-free model, the retrieved jet speed is $\sim$1 km/s higher than the GCM average, but this is a relatively small difference. On the other hand, the retrieved jet width comes out lower. This could, in part, be a geometric effect as the jet is defined as a Gaussian \emph{on the planet disk} in our forward model (with $\sigma_\text{jet}$ in units of $R_p$), while the Gaussian fit to the GCM jet was performed with latitude as a horizontal coordinate\footnote{To convert our retrieved jet width to degrees latitude in Fig. \ref{fig:1D_velocity_profiles}, we simply used $\sigma_\text{jet}[\text{deg}] = \arctan{(\sigma_\text{jet}[R_p])}$}. Even though our four-parameter velocity field is a radical simplification compared to the 3D wind profiles of UHJs, Fig. \ref{fig:1D_velocity_profiles} demonstrates that our retrievals are broadly able to recover the global wind properties of the GCM outputs.

The only wind parameter not discussed so far is $\varphi_\text{sink}$, the longitude towards which the source-to-sink flow converges. $\varphi_\text{sink}$ controls the projected velocities on the planet disk not via the norm of the wind vectors, but via their orientation. When $\varphi_\text{sink} = 180^\circ$, the flow is symmetric and the winds cross the terminator plane perpendicularly. However, from GCM studies of UHJs, we know that this description is not entirely correct. {\color{black}{That is, rotational effects organize the atmospheric circulation into an \emph{asymmetric} flow pattern and prevent simple convergence at the antistellar point.}} As a result, wind vectors tend to cross the terminator plane under different angles on the eastern hemisphere compared to the western hemisphere (e.g., \citealt{Tan2019,Wardenier2021,Beltz2022}), effectively converging to a (virtual) longitude that is offset from the antistellar point\footnote{The orthographic plots in Fig. 2 in \citet{Beltz2022} are a clear demonstration of this phenomenon for WASP-76b. Also, Fig. 5 in \citet{Tan2019} shows how the wind vectors of the source-to-sink flow have different orientations on the eastern hemisphere compared to the western hemisphere.}. In our retrievals, accounting for this offset turned out to be crucial to match the blueshifts of the GCM spectra during pre-eclipse. Fig. \ref{fig:corner_R1e5_all_phases} shows that we retrieve very strong directional asymmetries in the source-to-sink flow, with  $\varphi_\text{sink}$ ranging between $\sim$25$^\circ$ for the drag-free model and $\sim$75$^\circ$ for the strong-drag model. We recognize that these values may be harder to interpret physically, as a pure day-to-night flow would require $|\varphi_\text{sink}| >$ 90$^\circ$, but this is somewhat inherent to our simplified forward model, which only features four wind parameters.

Fig. \ref{fig:corner_R1e5_all_phases} also demonstrates that the parameters associated with the scale factor, $\alpha_0$ and $\phi_0$, are tightly constrained. For the models with drag, $\phi_0$ lies close to zero, indicating that the temperature structures are roughly symmetric about the substellar point. For the drag-free model, whose hotspot is shifted eastward, we retrieve a negative value for $\phi_0$, which implies that the emission lines are strongest post-eclipse. This is in agreement with the notion that the temperature gradients, which set the line strengths via $B_\nu(T_\text{line core}) - B_\nu(T_\text{continuum})$, are strongest on the western dayside, away from the hotspot (\citealt{vanSluijs2022,vansluijs2025,Wardenier2025}). {\color{black}{The reason we find $\alpha_0 < 1$ is somewhat arbitrary. It turns out that our 1D template has a slightly stronger line contrast than the 3D ``static'' dayside spectrum (with no dynamics) due to the different underlying temperature structures. As a result, we need a scale-factor amplitude that is slightly smaller than unity to correctly reproduce the line strengths. In short, $\phi_0$ should only depend on the ``data'', while $\alpha_0$ depends on both the ``data'' and the template used in the retrieval.}}

\begin{table}
\caption{Summary of the retrievals for pre- and post-eclipse at different resolutions.}
\vspace{-10pt}
\label{tab:priors_posteriors_resolution}
\centering
\small
\renewcommand{\arraystretch}{1.3}
\begin{tabular}{lcccc}
\hline\hline
\textbf{Param.} & \textbf{$R$/1000} & \textbf{No drag} & \textbf{Weak drag} & \textbf{Strong drag} \\
\hline

 & 500 & $8.1^{+0.3}_{-0.2}$ & $9.9^{+0.1}_{-0.6}$ & $3.6^{+0.1}_{-0.2}$ \\
$v_{\mathrm{jet}}$
 & 100 & $7.8^{+0.1}_{-0.6}$ & $6.3^{+0.9}_{-1.3}$ & $7.8^{+0.5}_{-0.4}$ \\
\ \ $[\mathrm{km/s}]$
 & 50 & $5.6^{+1.4}_{-2.1}$ & $5.7^{+3.4}_{-5.6}$ & $0.1^{+0.2}_{-0.1}$ \\
\hline

 & 500 & $0.25^{+0.01}_{-0.01}$ & $0.04^{+0.01}_{-0.01}$ & $0.12^{+0.01}_{-0.01}$ \\
$\sigma_{\mathrm{jet}}$
 & 100 & $0.23^{+0.02}_{-0.01}$ & $0.01^{+0.01}_{-0.01}$ & $0.04^{+0.01}_{-0.01}$ \\
\ \ $[{R}_p]$
 & 50 & $0.19^{+0.11}_{-0.03}$ & $0.01^{+0.23}_{-0.01}$ & $0.23^{+0.06}_{-0.14}$ \\
\hline

 & 500 & $3.2^{+0.5}_{-0.1}$ & $3.2^{+0.1}_{-0.1}$ & $1.6^{+0.1}_{-0.1}$ \\
$v_{\mathrm{wind}}$
 & 100 & $2.8^{+0.3}_{-0.1}$ & $2.9^{+0.1}_{-0.1}$ & $1.5^{+0.1}_{-0.1}$ \\
\ \ $[\mathrm{km/s}]$
 & 50 & $2.8^{+0.5}_{-0.3}$ & $3.1^{+0.5}_{-0.2}$ & $1.6^{+0.1}_{-0.1}$ \\
\hline

 & 500 & $25^{+1}_{-2}$ & $62^{+1}_{-2}$ & $75^{+1}_{-1}$ \\
$\varphi_{\mathrm{sink}}$
 & 100 & $28^{+9}_{-1}$ & $60^{+1}_{-1}$ & $76^{+2}_{-1}$ \\
\ \ $[\mathrm{deg}]$
 & 50 & $29^{+5}_{-4}$ & $59^{+4}_{-8}$ & $66^{+2}_{-1}$ \\
\hline

 & 500 & $0.18^{+0.16}_{-0.02}$ & $0.44^{+0.04}_{-0.02}$ & $0.39^{+0.01}_{-0.01}$ \\
$w_0$
 & 100 & $0.06^{+0.05}_{-0.01}$ & $0.32^{+0.02}_{-0.03}$ & $0.29^{+0.02}_{-0.02}$ \\
 & 50 & $0.04^{+0.07}_{-0.03}$ & $0.37^{+0.32}_{-0.11}$ & $0.22^{+0.05}_{-0.02}$ \\
\hline

 & 500 & $0.88^{+0.01}_{-0.01}$ & $0.91^{+0.01}_{-0.01}$ & $0.83^{+0.01}_{-0.01}$ \\
$\alpha_0$
 & 100 & $0.89^{+0.01}_{-0.01}$ & $0.91^{+0.01}_{-0.01}$ & $0.85^{+0.01}_{-0.01}$ \\
 & 50 & $0.89^{+0.01}_{-0.01}$ & $0.95^{+0.01}_{-0.01}$ & $0.88^{+0.01}_{-0.01}$ \\
\hline

 & 500 & $-0.032^{+0.001}_{-0.001}$ & $0.017^{+0.002}_{-0.002}$ & $0.000^{+0.001}_{-0.001}$ \\
$\phi_0$
 & 100 & $-0.034^{+0.002}_{-0.005}$ & $0.010^{+0.003}_{-0.002}$ & $-0.007^{+0.001}_{-0.001}$ \\
 & 50 & $-0.038^{+0.003}_{-0.003}$ & $0.001^{+0.004}_{-0.005}$ & $-0.012^{+0.003}_{-0.002}$ \\
\hline

\end{tabular}
\vspace{-6pt}
\tablefoot{Each row pertains to a free parameter in the forward model and shows the posterior values obtained for each GCM output (columns) at three different spectral resolutions $R$.}
\end{table}

\begin{table}
\caption{Summary of the retrievals at $R$ = 100,000 for different orbital phases.}
\vspace{-10pt}
\label{tab:priors_posteriors_phases}
\centering
\small
\renewcommand{\arraystretch}{1.3}
\setlength{\tabcolsep}{5pt}
\begin{tabular}{l c c c c}
\hline\hline
\textbf{Param.} & \textbf{Phases} & \textbf{No drag} & \textbf{Weak drag} & \textbf{Strong drag} \\
\hline

 & all & $7.8^{+0.1}_{-0.6}$ & $6.3^{+0.9}_{-1.3}$ & $7.8^{+0.5}_{-0.4}$ \\
$v_{\mathrm{jet}}$ & pre & $8.8^{+0.5}_{-0.3}$ & $7.4^{+1.2}_{-7.3}$ & $8.8^{+0.8}_{-1.4}$ \\
\ \ $[\mathrm{km/s}]$ & post & $7.3^{+0.2}_{-0.2}$ & $7.5^{+0.5}_{-0.5}$ & $4.3^{+1.9}_{-1.0}$ \\
\hline

 & all & $0.23^{+0.02}_{-0.01}$ & $0.01^{+0.01}_{-0.01}$ & $0.04^{+0.01}_{-0.01}$ \\
$\sigma_{\mathrm{jet}}$ & pre & $0.19^{+0.01}_{-0.01}$ & $0.01^{+0.27}_{-0.01}$ & $0.02^{+0.02}_{-0.01}$ \\
\ \ $[R_p]$ & post & $0.26^{+0.01}_{-0.01}$ & $0.08^{+0.01}_{-0.01}$ & $0.08^{+0.01}_{-0.01}$ \\
\hline

 & all & $2.8^{+0.3}_{-0.1}$ & $2.9^{+0.1}_{-0.1}$ & $1.5^{+0.1}_{-0.1}$ \\
$v_{\mathrm{wind}}$ & pre & $3.2^{+0.3}_{-0.1}$ & $2.7^{+0.2}_{-0.2}$ & $1.3^{+0.2}_{-0.1}$ \\
\ \ $[\mathrm{km/s}]$ & post & $2.2^{+0.1}_{-0.1}$ & $3.4^{+0.1}_{-0.1}$ & $1.7^{+0.1}_{-0.1}$ \\
\hline

 & all & $28^{+9}_{-1}$ & $60^{+1}_{-1}$ & $76^{+2}_{-1}$ \\
$\varphi_{\mathrm{sink}}$ & pre & $47^{+2}_{-1}$ & $67^{+5}_{-4}$ & $87^{+3}_{-4}$ \\
\ \ $[\mathrm{deg}]$ & post & $17^{+5}_{-4}$ & $33^{+5}_{-2}$ & $51^{+3}_{-3}$ \\
\hline

 & all & $0.06^{+0.05}_{-0.01}$ & $0.32^{+0.02}_{-0.03}$ & $0.29^{+0.02}_{-0.02}$ \\
$w_0$ & pre & $0.01^{+0.05}_{-0.01}$ & $0.23^{+0.19}_{-0.06}$ & $0.20^{+0.14}_{-0.02}$ \\
 & post & $0.00^{+0.01}_{-0.00}$ & $0.75^{+0.07}_{-0.09}$ & $0.38^{+0.06}_{-0.04}$ \\
\hline

 & all & $0.89^{+0.01}_{-0.01}$ & $0.91^{+0.01}_{-0.01}$ & $0.85^{+0.01}_{-0.01}$ \\
$\alpha_0$ & pre & $0.94^{+0.02}_{-0.01}$ & $0.95^{+0.01}_{-0.02}$ & $0.89^{+0.01}_{-0.01}$ \\
 & post & $1.17^{+0.01}_{-0.02}$ & $1.10^{+0.02}_{-0.02}$ & $0.97^{+0.01}_{-0.01}$ \\
\hline

 & all & $-0.034^{+0.002}_{-0.005}$ & $0.010^{+0.003}_{-0.002}$ & $-0.007^{+0.001}_{-0.001}$ \\
$\phi_0$ & pre & $-0.073^{+0.006}_{-0.005}$ & $-0.023^{+0.017}_{-0.009}$ & $-0.040^{+0.010}_{-0.005}$ \\
 & post & $0.094^{+0.004}_{-0.004}$ & $0.081^{+0.005}_{-0.005}$ & $0.061^{+0.004}_{-0.004}$ \\
\hline

\end{tabular}
\vspace{-6pt}
\tablefoot{Rows show the posterior values obtained for each GCM output (columns) across pre- and post-eclipse (``all''), pre-eclipse only (``pre''), and post-eclipse only (``post'').}
\end{table}

\begin{figure*}
    \centering
    \vspace{20pt}

    \begin{minipage}[t]{0.485\textwidth}
        \centering
        \includegraphics[width=\textwidth]{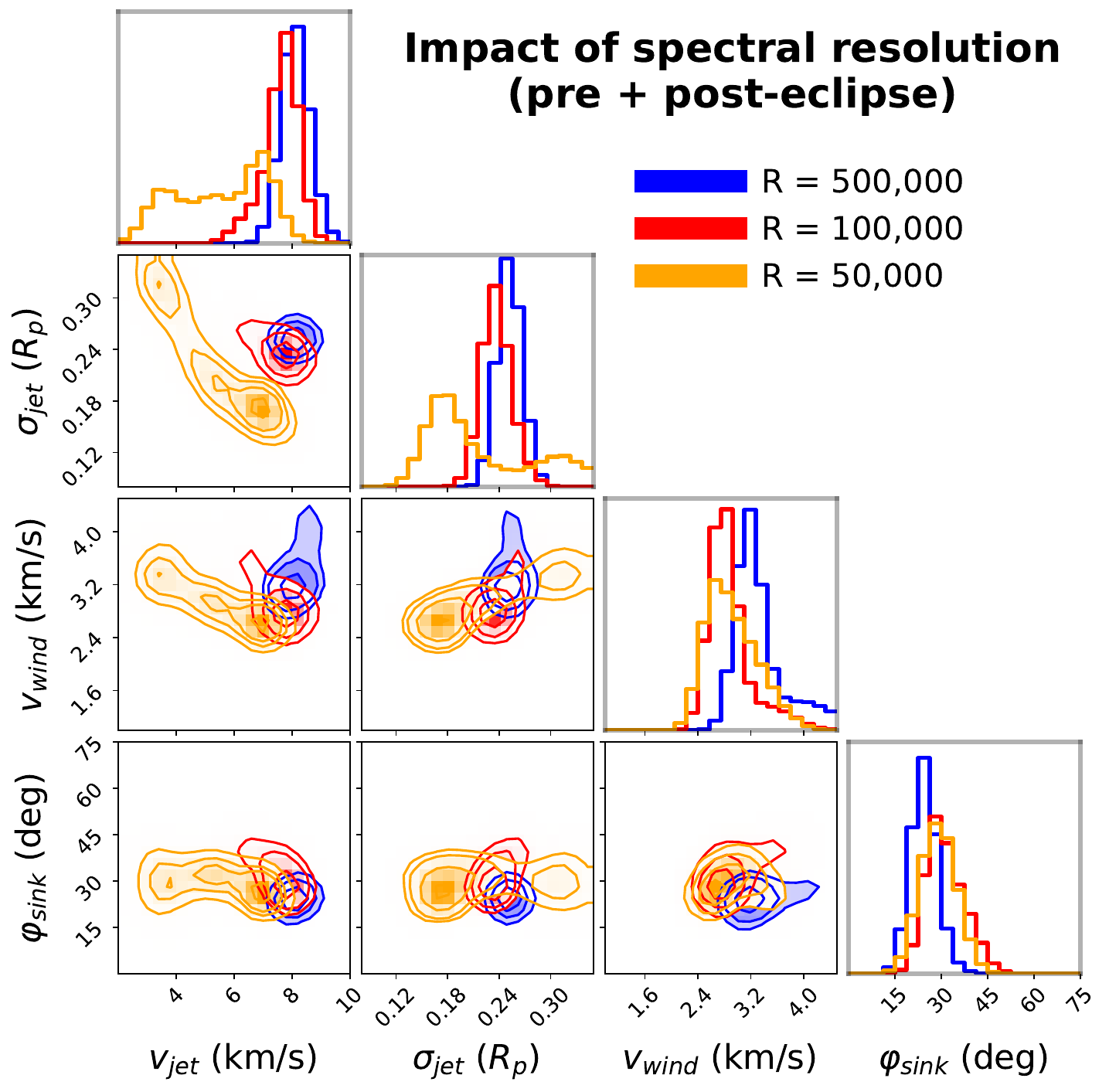}
    \end{minipage}
    \hfill
    \begin{minipage}[t]{0.485\textwidth}
        \centering
        \includegraphics[width=\textwidth]{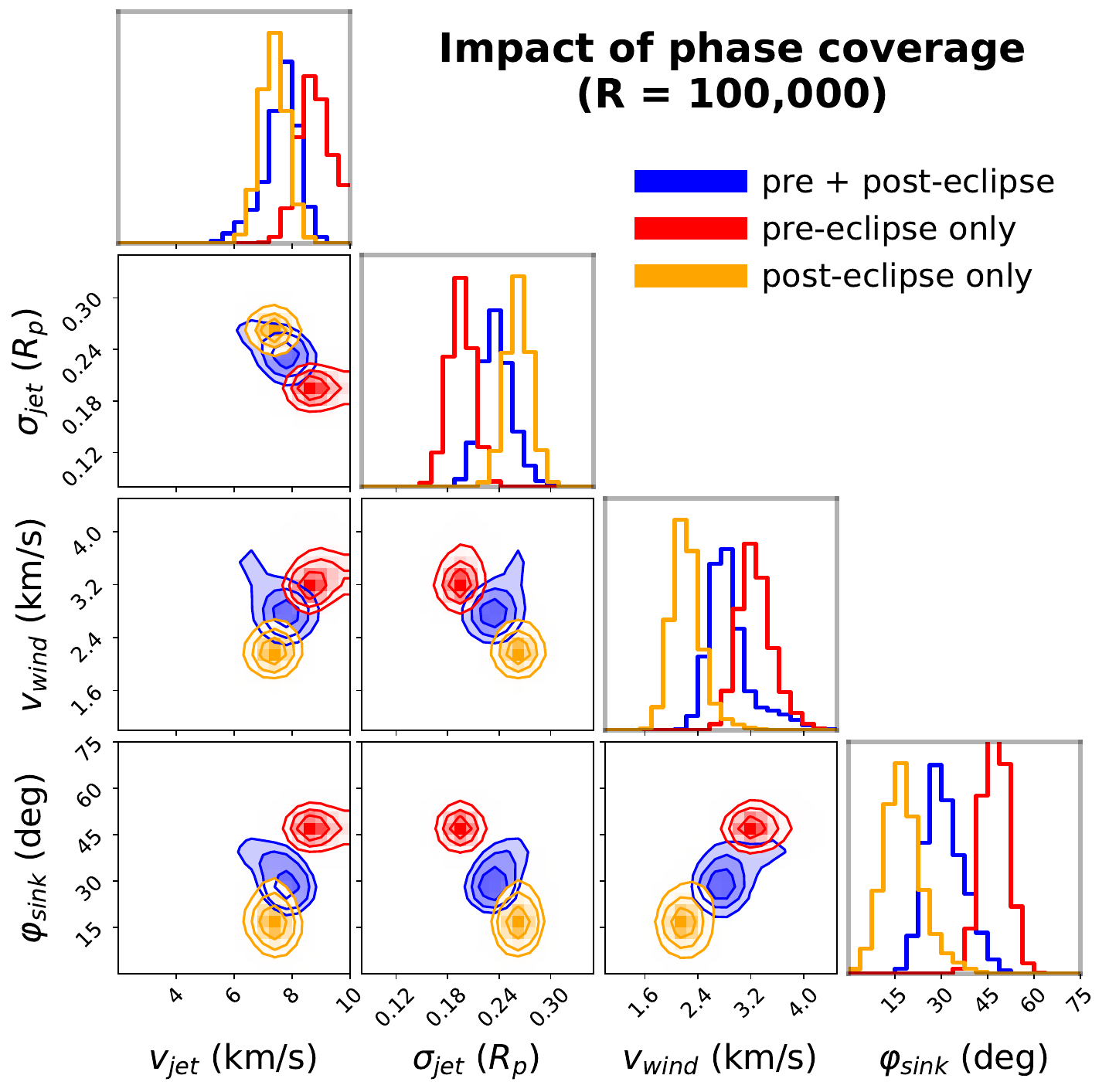}
    \end{minipage}

    \vspace{-3pt}
    \caption{Corner plots showing the impact of spectral resolution (left panel) and the impact of orbital-phase coverage (right panel) on the posterior distributions of the four wind-field parameters obtained for the \emph{drag-free} GCM output of WASP-76b. The full posteriors for all parameters included in the retrieval can be found in Appendix \ref{app:B}.}

    \vspace{0pt}
    
    \label{fig:two_panel_fullwidth}
\end{figure*}

Our last physical parameter is the disk-center weight $w_0$, which controls how the velocity field is weighted when we compute the broadening kernel (the weight at the disk edge is fixed at 1). As shown in Fig. \ref{fig:corner_R1e5_all_phases}, the retrievals all prefer a more \emph{limb-weighted} scenario, where the weights assigned to the projected velocities increase towards the edge of the planet disk\footnote{This is the reason we did not write the weight function in terms of the typical $u_1$ coefficient used in limb-darkening equations, as $w_0\rightarrow0$ would imply $u_1\rightarrow\infty$.}. In physical terms, this means that light rays emerging from the limb encounter a larger line-of-sight temperature gradient than light rays emerging from the center. Because our GCM outputs feature strong vertical temperature inversions that reach down to pressures of 0.1 mbars at some longitudes (Fig. 3 in \citealt{Wardenier2025}), this is not an unreasonable result. {\color{black}{Near the limb, the pressures probed by a line core and its surrounding continuum shift to lower pressures due to the viewing geometry, and this could result in a stronger temperature gradient being probed. It is important to stress, however, that fitting a single value of $w_0$ to \emph{all} phases is a strongly simplified representation of the true underlying 3D temperature structure.}} In Section \ref{sec:advanced_weight_function}, we will explore more physically motivated weight functions, but these unavoidably come with a higher model complexity.

\subsection{Impact of spectral resolution and phase coverage}
\label{ref:impact_reso_phases}

In this section, we discuss the impact of spectral resolution and phase coverage on the inferred broadening-kernel parameters. Tables \ref{tab:priors_posteriors_resolution} and \ref{tab:priors_posteriors_phases} provide a complete overview of the posterior values found from each retrieval. Fig. \ref{fig:two_panel_fullwidth} specifically focuses on the drag-free model and shows corner plots for the wind-field parameters at different spectral resolutions (keeping the phase coverage fixed) and different phase coverages (keeping the spectral resolution fixed), respectively.

As shown in the left corner plot in Fig. \ref{fig:two_panel_fullwidth}, the retrievals are mostly consistent across resolutions in terms of the source-to-sink flow, with $v_{\mathrm{wind}}$ being slightly higher at $R$ = 500,000. This is reassuring: the (asymmetric) source-to-sink flow mostly impacts the phase-dependent line \emph{shifts}. Because we do not consider any noise in our retrievals, the line positions remain well-defined, even at lower resolutions. The story is different for the equatorial jet. While $v_{\mathrm{jet}}$ and $\sigma_{\mathrm{jet}}$ are consistent between the higher-resolution retrievals, the retrieval at $R$ = 50,000 struggles to constrain the jet. This is not surprising: the jet and the instrument profile (assumed to be Gaussian in our case) both have a \emph{broadening} effect on the spectral lines. Therefore, once the spectral resolution becomes too low, the instrument profile can start mimicking the Doppler-broadening due to the jet.

Another interesting observation is the ``spurious'' inference of a narrow jet (with $v_{\mathrm{jet}}$ = 3.6 km/s and $\sigma_{\mathrm{jet}}$ = 0.12 $R_{p}$) in the strong-drag model at $R$ = 500,000 (see Table \ref{tab:priors_posteriors_resolution}). As we discussed previously, this is likely related to the simplified nature of our model: at extremely high resolution, our two-parameter source-to-sink flow cannot fully account for the line shapes in the GCM spectra, so the retrieval compensates for this by including a jet. In this respect, a resolution of $\sim$100,000 may constitute the perfect trade-off: high enough to resolve the global features of the 3D wind profile, but low enough to ``blur'' out the effects of the more complicated features that are not adequately captured by our simple velocity-field parametrization. In real observations, there will also be some additional blurring due to the planet orbital motion during each exposure.

\begin{figure*}[t!]
        \centering
        \vspace{0pt}
        \makebox[\textwidth][c]{\hspace{-10pt}\includegraphics[width=1.1\textwidth]{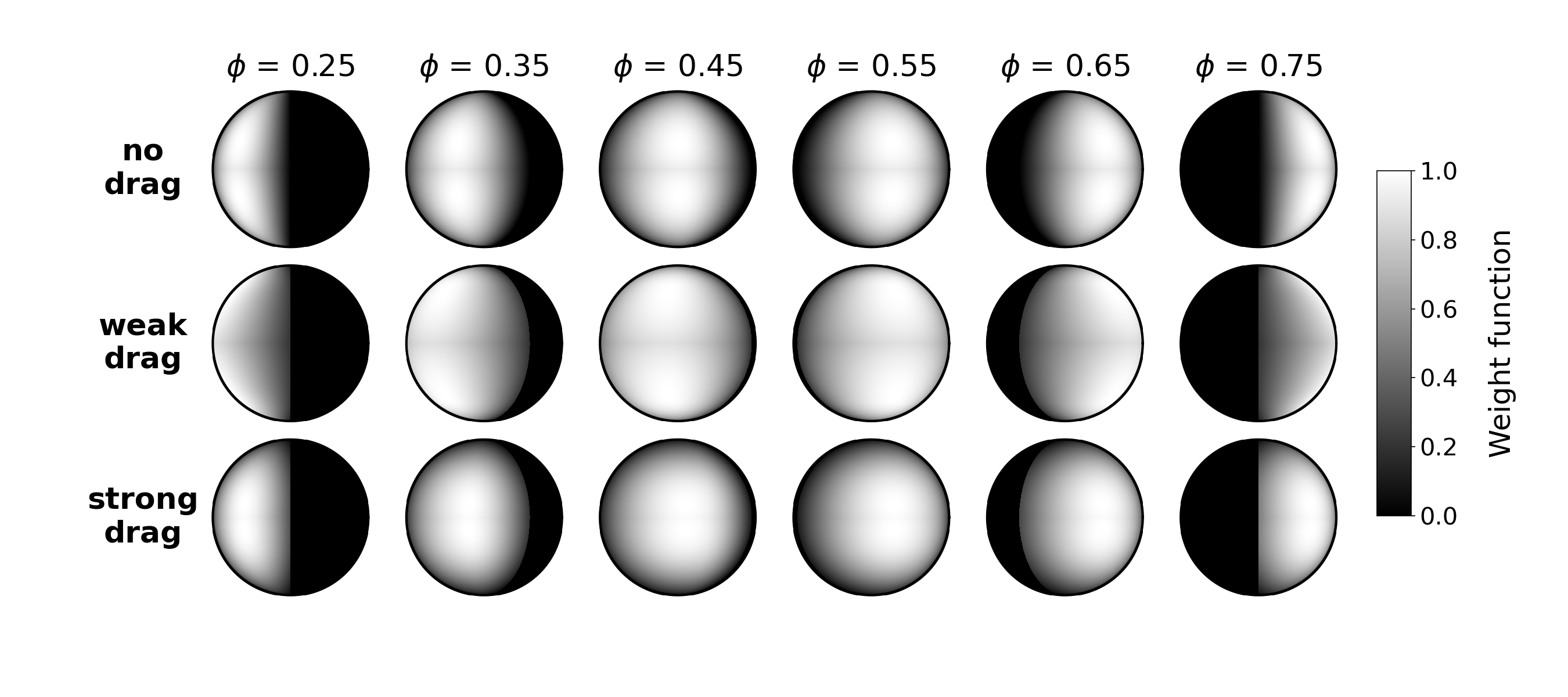}}
        \vspace{-44pt}
        \caption{Best-fit weight functions for the $R$ = 100,000 retrievals covering both pre-eclipse and post-eclipse phases. Each row depicts the result for a different GCM output of WASP-76b. At each phase $\phi$, the nightside is masked out as our retrievals only recover information from the dayside of the planet.}
        \label{fig:best_fit_weight_masks}
\end{figure*}

\begin{figure}
       \centering
        \vspace{0pt}
        \includegraphics[width=0.49\textwidth]{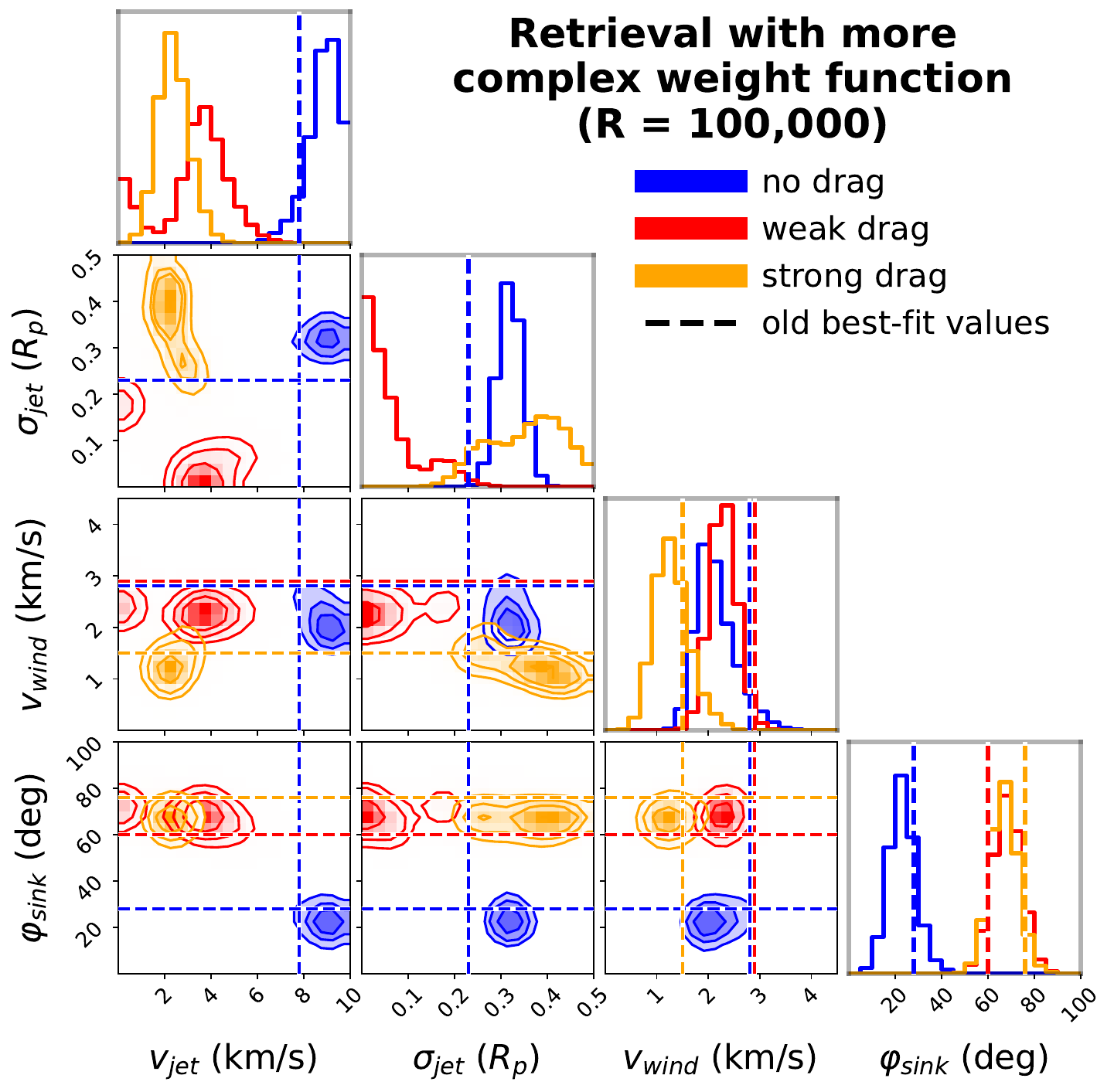}
        \vspace{-10pt}
        \caption{Corner plots of the wind-field parameters obtained from the retrievals with the more complex weight function described in Section \ref{sec:advanced_weight_function}, covering both pre-eclipse and post-eclipse (at \mbox{$R=$ 100,000}). The dashed lines show the best-fit values from the nominal (``old'') retrievals in which $w_0$ was the only parameter describing the weight function.}
        \label{fig:corner_hotspot_offset}
\end{figure}

\begin{figure*}[t!]
        \centering
        \vspace{-20pt}
        \makebox[\textwidth][c]{\hspace{-10pt}\includegraphics[width=1.11\textwidth]{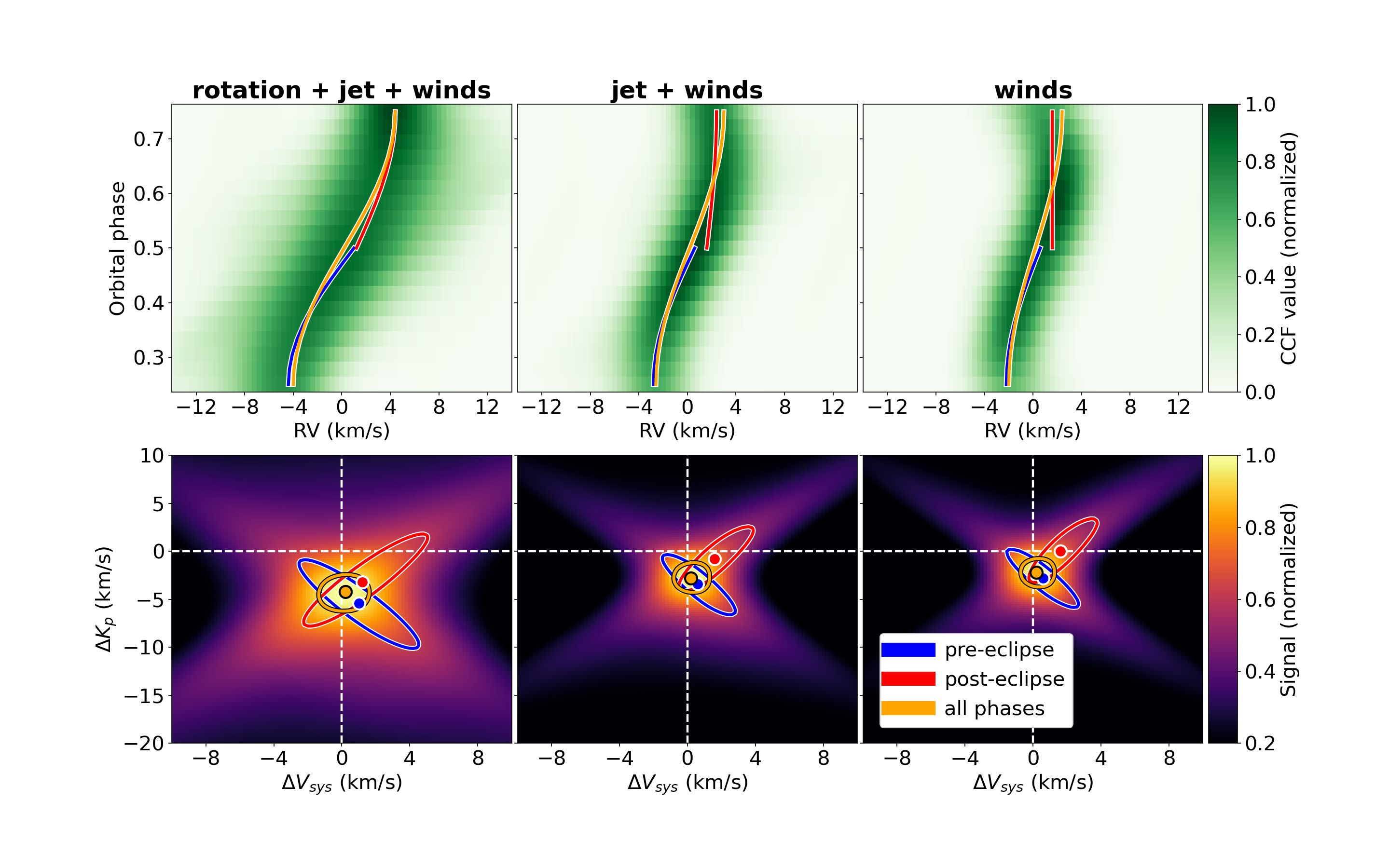}}
        \vspace{-44pt}
        \caption{CCF maps in the planet rest frame (top row) and associated $K_\text{p}$--$V_\text{sys}$ maps (bottom row) obtained by cross-correlating the best-fit spectra corresponding to the \emph{drag-free} GCM with the original 1D template. \textbf{Top row:} From left to right, the panels illustrate how the CCF trail changes when successively switching off contributions from planet rotation and the equatorial jet. The right panel shows the CCF trail when only accounting for the source-to-sink flow (with $v_\text{wind}$ = 2.8 km/s and $\varphi_\text{sink}$ = 28$^\circ$). In each of the panels, we plot the sinusoids that correspond to the best-fit $\Delta K_\text{p}$ and $\Delta V_\text{sys}$ values in pre-eclipse only (blue), in post-eclipse only (red), and across all phases (yellow), respectively. \textbf{Bottom row:} $K_\text{p}$--$V_\text{sys}$ maps corresponding to the combined pre- and post-eclipse phases. On top of each map, we plot the best-fit ($\Delta K_\text{p}$, $\Delta V_\text{sys}$) values associated with the sinusoids in the top row (circular markers). The contour associated with each marker indicates the region where the integrated CCF signal is within 10$\%$ of the maximum.}
        \label{fig:ccf_and_kp_vsys_maps}
\end{figure*}

The right corner plot in Fig. \ref{fig:two_panel_fullwidth} shows the retrievals for the drag-free model at $R$ = 100,000 for different phase coverages. Qualitatively, the results are consistent: we also retrieve the equatorial jet solely based on the pre-eclipse or post-eclipse phases. For each parameter, the posterior from the retrieval covering all phases lies between the posterior distributions from the separate pre- and post-eclipse retrievals. The corner plot also reveals some correlations between the different parameters \emph{across} retrievals. For example, the jet width correlates inversely with the day-to-night wind speed, showing that the jet can take over some Doppler effects from the source-to-sink flow (and vice versa) when a more limited phase range is considered. The same holds for $v_{\mathrm{wind}}$ and $\varphi_{\mathrm{sink}}$: a stronger directional asymmetry in the source-to-sink flow requires lower wind speeds to match the observed Doppler shifts. 

In general, the larger the range of phases sampled, the better the constraint on the wind-field orientation, and the more reliable the inferred absolute wind speeds. Of course, one assumption that we make here is that the wind speeds on the western dayside are not too different from those on the eastern dayside. To verify whether this is indeed the case for our GCM outputs, we computed 1D averages similar to those in Fig. \ref{fig:1D_velocity_profiles}. We found negligible differences between the average wind speeds on both hemispheres, which cannot explain the $\sim$1 km/s discrepancy between the $v_{\mathrm{wind}}$ values inferred from the separate pre- and post-eclipse retrievals in Fig. \ref{fig:two_panel_fullwidth}.

\subsection{More complex weight functions}
\label{sec:advanced_weight_function}

Our nominal broadening-kernel model features only one parameter, $w_0$, that controls how the velocity field is weighted when constructing the Doppler kernel. In reality, however, the line strengths associated with a certain position on the planet disk are set by the probed temperature gradient via $B_\nu(T_\text{line core}) - B_\nu(T_\text{continuum})$. In this section, we will explore a weight function that aims to more realistically mimic such a ``temperature-gradient map''. {\color{black}{In what follows, we will explicitly avoid using the term ``hotspot'' as it tends to be associated with absolute flux.}} 

We introduce three new parameters: a longitudinal peak offset $\varphi_\text{peak}$, a latitudinal offset $\theta_\text{peak}$ that splits the peak in two (such that the maxima lie at latitudes $\pm\theta_\text{peak}$), and a dropoff parameter $f_\text{d}$ that sets the weight of the terminator compared to the peak values, which always have a weight of 1. Between the weight peak(s) and the terminator, we assume that the weight function drops off sinusoidally from 1 to ($1 - f_\text{d}$). In this way, $f_\text{d}$ = 1 implies a maximally concentrated weight peak, while $f_\text{d}$ = 0 would produce a uniformly weighted dayside. We also keep the $w_0$ parameter to account for center-to-limb variations in the weight function. {\color{black}{We refer to Appendix \ref{app:A_1} for the full set of mathematical equations behind this more complex weight function.}} For the new free parameters, we use the following priors: $\mathcal{U}(-80,80)$ for $\varphi_\text{peak}$ [deg], $\mathcal{U}(0,80)$ for $\theta_\text{peak}$ [deg], and $\mathcal{U}(0,1)$ for $f_\text{d}$, where $\mathcal{U}$ denotes a uniform distribution. 

Fig. \ref{fig:best_fit_weight_masks} shows the new weight functions retrieved for each of the three GCM outputs. We reiterate that these should, in principle, measure the temperature gradient rather than the absolute temperature. {\color{black}{The plots show that all retrievals prefer the inclusion of a weight peak,}} such that the weights drop off towards the terminator region (implying nonzero $f_\text{d}$). Also, the retrievals for the drag-free and the weak-drag models prefer a latitudinal offset, whereby the regions with the largest weights lie north and south of the equator. 

Fig. \ref{fig:corner_hotspot_offset} shows the inferred values of the wind-field parameters when accounting for the new weight function. With respect to the nominal retrievals (for which the best fits are marked by the dashed lines), there are a few differences to point out. First, the jet speed found for the drag-free model is $\sim$1 km/s higher than before. This agrees less well with the jet speed from the GCM plotted in Fig. \ref{fig:1D_velocity_profiles}. Second, the retrieval for the strong-drag model prefers a broad jet feature with $v_\text{jet} \sim$2.5 km/s, which is not present in the GCM output. Third, the inferred day-to-night wind speeds are 0.5-1 km/s lower for all three models compared to the retrievals with only one parameter describing the weight function. The full corner plots for the complex-weight-function retrievals can be found in Appendix \ref{app:B}.

From a more abstract standpoint, it is not completely surprising that we encounter a degeneracy between the velocity field and the weight function. The broadening kernel, whose shape we ultimately retrieve, is a product of the unweighted velocity histogram times a weight function, after which it is normalized so that the integrated surface area is unity. Therefore, two different sets of velocity distributions and weight functions could, in theory, give rise to the same broadening kernel. Subsequently, an Occam's razor-like test would give preference to the model with the smallest number of parameters (unless there is a strongly motivated prior). All in all, there is no single correct way to account for the temperature structure of the atmosphere: a model with too few parameters may not be able to fully capture the fine details of the line shapes, while too many parameters make a model prone to overfitting. Based on the retrievals in this work, we argue that a simple linear weight prescription (with disk-center weight $w_0$) is sufficient to correctly capture the global wind properties of an UHJ atmosphere. When comparing a model to real observational data, it remains important to perform a Bayesian evidence-test to assess whether adding an extra parameter is justified. 

\subsection{$K_\text{p}$ offsets from to day-to-night winds}
\label{sec:delta_kp_offset} 

A useful feature of our broadening-kernel model is that we can easily ``switch off'' different contributions in the velocity field and assess the impact on the $\Delta K_\text{p}$ and $\Delta V_\text{sys}$ values associated with the spectral time series (e.g., \citealt{Wardenier2023,Wardenier2025}). Fig. \ref{fig:ccf_and_kp_vsys_maps} shows the result of such an exercise. We take the best-fit spectra associated with the drag-free GCM output (see \mbox{Fig. \ref{fig:best_fit_spectra}}) and cross-correlate them with the original 1D emission template to obtain a 2D cross-correlation function (CCF). As shown in the left column of Fig. \ref{fig:ccf_and_kp_vsys_maps}, we first use the broadening kernel that accounts for planet rotation, the equatorial jet, and the source-to-sink flow. We then successively set $v_\text{rot}$ and $v_\text{jet}$ to zero, such that only the contribution from the source-to-sink flow remains (see right column). We compute the $K_\text{p}$--$V_\text{sys}$ maps by co-adding the CCF values along sinusoids of the form $v(\phi) = \Delta V_\text{sys} + \Delta K_\text{p} \sin(2\pi\phi)$ -- see \citet{Wardenier2025} for further details.

There are two main lessons to be learned from Fig. \ref{fig:ccf_and_kp_vsys_maps}. The first, more general lesson is that the CCF trail in the planet rest frame will never be a perfect sinusoid with a period $P$ (where $P$ is the orbital period of the planet). Therefore, one may find \emph{different} $\Delta K_\text{p}$ and $\Delta V_\text{sys}$ values for \emph{the same} CCF trail, depending on the orbital-phase range that is considered. The right panel illustrates this particularly well. While the pre-eclipse part of the trail is best described by a curve with a slope (i.e., a sinusoid with nonzero amplitude), the post-eclipse part is best described by a vertical line (i.e., a sinusoid with zero amplitude). This results in discrepant $\Delta K_\text{p}$ and $\Delta V_\text{sys}$ values, even though they were calculated from the same CCF map (see also Fig. 13 in \citealt{Wardenier2025}).

A second lesson is that the source-to-sink flow can induce both $K_\text{p}$ and $V_\text{sys}$ offsets. In fact, Fig. \ref{fig:ccf_and_kp_vsys_maps} suggests that the source-to-sink flow is a stronger driver of the phase-dependent Doppler shifts than the equatorial jet. This may not be surprising, given that the source-to-sink flow covers a much larger area on the planet disk than the equatorial jet (see Fig. \ref{fig:doppler_components}). Also, because $\varphi_\text{sink}$ is strongly offset from the antistellar point in the best-fit model, the net Doppler shifts induced by the source-to-sink flow are no longer symmetric about phase 0.5. This means that we need a nonzero $\Delta K_\text{p}$ to fit the CCF trail across pre- and post-eclipse simultaneously (see the yellow curve in the top-right panel). In contrast, for a symmetric source-to-sink flow (with $\varphi_\text{sink} = \pm 180^\circ$), the net Doppler shift at phases $0.5 + x$ and $0.5 - x$ should be the same (ignoring all other contributions). In Appendix \ref{app:B}, we further explore the impact of the sink longitude on the shape of the CCF trail and the resulting $\Delta K_\text{p}$ and $\Delta V_\text{sys}$ values. 

Finally, in \citet{Wardenier2025}, we postulated that the $K_\text{p}$ offset of a planet should be capped by its equatorial rotation velocity and jet speed: \mbox{$|\Delta K_{\text{p}}| < (v_\text{eq} + v_\text{jet})$}. However, our current work shows that part of the observed $\Delta K_\text{p}$ can also be due to the source-to-sink flow. Therefore, not all offsets in excess of the rotation velocity should automatically be attributed to the possible presence of a jet. This ``qualitative'' degeneracy underscores the need for a broadening-kernel model that directly maps the observed Doppler shifts onto line-of-sight velocities on the planet disk.

\subsection{Accounting for the nightside contribution}
\label{sec:accounting_for_nightside}

In this work, we focused only on retrieving the velocity fields on the daysides of UHJs. With the current generation of telescopes, this is likely the best we can do, as nightsides are much harder to detect via their absorption features, let alone characterize. To the knowledge of the authors, WASP-33b is the only UHJ for which multiple studies have claimed the detection of nightside flux at high resolution. These were based on in-transit observations of the planet (\citealt{Yang2024}), and post-transit observations with $\phi < 0.16$ (\citealt{Mraz2024}).

It would be conceptually simple, however, to extend the retrieval to also fit the nightside part of the spectrum (see also \citealt{Zhang2026}). To this end, we would start with two 1D templates: one containing only emission features and the other containing only absorption features. We would then construct two broadening kernels (optionally with different sets of parameters), whose weights are zero on the nightside and the dayside, respectively. We then convolve both templates with their respective kernels, multiply them by a (phase-dependent) scale factor, and add the results together to obtain the total emission spectrum.

For cooler gas giants that do not feature thermal inversions on their daysides, it could be sufficient to use only a single 1D template containing absorption features. {\color{black}{We should note, however, that the temperature structure alone does not set the line shapes. Differences in chemical abundances between the dayside and the nightside would also result in more nuanced changes in the spectrum.}}

\subsection{Comparison to Zhang et al. (2026)}

In this section, we compare our broadening-kernel model to the forward model presented by \citet{Zhang2026}, who retrieved the dynamical properties of the UHJ KELT-9b based on post-transit and pre-eclipse observations from KPF on Keck. \citet{Zhang2026} used four free parameters to describe Doppler effects in the spectrum: the orbital velocity $K_\text{p}$, a systemic velocity offset $\Delta V_\text{sys}$, a (day-to-night) wind speed $u_\varphi$, and the full width at half maximum (FWHM) of the spectral lines at eclipse. The authors assumed an analytical velocity profile on the planet disk, given by

\begin{equation} \label{eq:line_doppler}
    v_\mathrm{LOS}(\varphi, \theta) = (v_\mathrm{rot} + u_\varphi\sin\varphi) \sin(\phi+\varphi) \cos\theta,
\end{equation}

\noindent where $v_\mathrm{LOS}(\varphi, \theta)$ is the line-of-sight velocity at longitude $\varphi$ and latitude $\theta$, and $v_\mathrm{rot}$ is the rotational velocity at the equator. The effective Doppler shift at a certain orbital phase $\phi$ (say, associated with the dayside) was then computed as follows: (i) sample the planet disk on a grid, (ii) evaluate Eq. (\ref{eq:line_doppler}) at all visible latitudes and longitudes on the given hemisphere, (iii) Doppler shift a single-line toy model in each grid cell according to the local $v_\mathrm{LOS}$, and (iv) add all the shifted lines together, weighted by the projected area of their corresponding cell. The net Doppler shift is given by the peak position of the resulting flux. 

Conceptually, the approach from \citet{Zhang2026} is very similar to our work, as it tries to directly link Doppler shifts in the spectrum to line-of-sight velocities on the planet disk. When it comes to the source-to-sink flow, \citet{Zhang2026} assumed that the wind speeds increase monotonically with longitude, from zero at the substellar point to $u_\varphi$ at the terminator. This different choice of parametrization would be straightforward to include in our forward model, too. On the other hand, Eq. (\ref{eq:line_doppler}) does not account for a ``sink'' that is offset from the antistellar point, which allows for more flexibility but also makes the math more complicated (see Appendix \ref{app:A_0}). 

\citet{Zhang2026} noticed that the line broadening obtained from their co-added toy models did not match the line broadening observed in the KELT-9b emission spectra. To mitigate this, they used a separate parametrization for the phase-dependent line widths, given by FWHM = $\omega \sin^2(\pi \phi)$, where $\omega$ is the maximum line width at eclipse. This is different from our parametrization, which fits both the line shifts and the line widths in a self-consistent manner. 

Another thing to note is that \citet{Zhang2026} simultaneously retrieved the vertical temperature profile when fitting their models to the data. For this reason, they set the amplitude of the scale factor (called $\alpha_0$ in our model) to unity. Retrieving the vertical temperature gradient and a scale factor at the same time would lead to degeneracies: shallower temperature gradients can be compensated for by a larger scale factor and vice versa. Because we use a single, pre-computed 1D template in our work, retrieving a scale factor is necessary to match the Doppler-broadened model to the GCM spectra.

\section{Summary and conclusion}
\label{sec:conclusion}

In this work, {\color{black}{we presented \texttt{dopplerkernel}}}\footnote{The Python code and a Jupyter Notebook tutorial are available on \href{https://github.com/joostwardenier/dopplerkernel}{\texttt{github.com/joostwardenier/dopplerkernel}}}, a forward model to generate broadening kernels that describe the impact of rotation and atmospheric dynamics on the thermal emission spectra of ultra-hot Jupiters (UHJs). Inspired by earlier work from \citet{Lesjak2024}, the model relies on parameterizing the velocity field on the planet disk, sampling this field on a grid, and building a histogram of weighted line-of-sight velocities to obtain the corresponding broadening kernel. The model is fast to evaluate, making it very suitable for implementation in a retrieval setup. 

Because our broadening-kernel model encompasses only a few parameters, it unavoidably comes with a number of simplifications. Motivated by recent models and observations (\citealt{Wardenier2025,Cont2025,Zhang2026}), one assumption we make is that the Doppler signatures associated with the nightside are negligible during the entirety of pre- and post-eclipse. Another simplification in our forward model is the absence of pressure-dependence. That is, the velocity field only varies as a function of latitude and longitude. Because the lines in an emission spectrum have different strengths, they should probe different temperatures and thus different pressures. This does not mean, however, that our model cannot be used to infer pressure-dependent wind information from observational data. In practice, one would partition the spectrum into subsets of emission lines with similar strength (e.g., \citealt{Miller-RicciKempton2012,Kesseli2024}) and perform separate retrievals for each subset, provided that the signal-to-noise ratio is good enough. This would allow one to resolve the wind field along the pressure axis.

Finally, our model aims to capture the full 3D wind profile of the planet in essentially four parameters: the jet speed $v_\text{jet}$, the jet width $\sigma_\text{jet}$, the source-to-sink flow speed $v_\text{wind}$, and a parameter $\varphi_\text{sink}$ that governs the orientation of the source-to-sink flow. The impact of the 3D temperature profile and the phase-dependent viewing geometry is described by three parameters: a parameter $w_0$ to account for the effect of limb-weighting, a scale factor amplitude $\alpha_0$, and a scale factor offset $\phi_0$. While our retrievals demonstrate that our broadening-kernel model is able to accurately reproduce the line shifts, shapes, and strengths in the emission spectra of our WASP-76b GCMs (see Fig. \ref{fig:best_fit_spectra}), it is important to realize that, ultimately, a model with seven free parameters can never fully do justice to the complexity of a 3D atmosphere. One could even ask the philosophical question whether there exists such a thing as \emph{the} wind speed of an UHJ (as suggested by our $v_\text{wind}$ parameter), given that the underlying velocity field is much more complicated.

In this work, we performed $K$-band retrievals on three GCM outputs of WASP-76b: a drag-free model, a weak-drag model, and a strong-drag model (see Fig. \ref{fig:zonal_winds}). Our retrievals successfully recovered the equatorial jet in the drag-free model at $\lesssim$ 1 km/s from the zonal mean. The jet could also be recovered when only considering pre- or post-eclipse phases. For each of the three models, we inferred source-to-sink flow speeds that are in good agreement with the average GCM wind speeds between 1 and 10 mbar (see Fig. \ref{fig:1D_velocity_profiles}). We also showed that separate pre- and post eclipse retrievals can return slightly different values for the source-to-sink flow speed (up to $\sim$1 km/s discrepant for the drag-free model) because of the directional asymmetries in the GCM wind profile. Sampling a larger range of orbital phases allows for more reliable constraints.

Furthermore, we conclude that spectral resolutions on the order of $R$ = 100,000 are suitable for these kinds of studies -- high enough to resolve the \emph{global features} in the wind profiles of UHJs (the jet and the source-to-sink flow), but also low enough to smear out more detailed features that a simplified parametric model inherently cannot reproduce. 

Finally, we experimented with more complex weight functions that account for the 3D temperature structure of the planet in a more physically motivated way. However, we observed that the addition of these extra parameters did not improve the inference of the jet speed in the drag-free model, while causing the retrieval to prefer the inclusion of a ``spurious'' ($\sim$2.5 km/s) jet for the strong-drag model. Again, we stress that the construction of a Doppler-broadening kernel is a fundamentally degenerate problem, as the kernel is the product of a velocity distribution and a weight function that is governed by the planet's 3D thermochemical structure, after which it is normalized. Adding too many parameters to a model can bring these degeneracies to the surface.

{\color{black}{Combining our \texttt{dopplerkernel} package with existing retrieval frameworks would allow us to constrain the chemical composition, temperature structure, and the wind profile of UHJs all at once. At the time of writing, there are plenty of high-quality, high-resolution emission datasets available to which our model can be applied, such as those of WASP-76b (ESPRESSO; \citealt{costasilva2024,Guilluy2025}),  WASP-121b (ESPRESSO, CRIRES+, and NIRPS; \citealt{Hoeijmakers2022,Pelletier2025,Smith2024,Bazinet2025}), WASP-189b (GHOST, CRIRES+, and IGRINS; \citealt{Deibert2024,Lesjak2024,Sanchez2025}), and KELT-9b (KPF; \citealt{Zhang2026}), among many others. Applying \texttt{dopplerkernel} to a suite of targets would enable a population-level study of atmospheric dynamics in emission, similar to work by \citet{Seidel2026} in transmission.}}   

With a new class of extremely large telescopes on the horizon, it is important to move away from purely 1D frameworks to analyze high-resolution observations of exoplanet atmospheres. Our broadening-kernel model is an attempt to account for phase-dependent Doppler signatures in the thermal emission spectra of UHJs (which are a product of both the 3D wind profile and the 3D thermochemical structure) while keeping computation time tractable and ensuring interpretability.

\begin{acknowledgements}
      We thank the referee for valuable comments that helped improve the quality of this manuscript. J.P.W. acknowledges support from the Swiss National Science Foundation (SNSF) under grant 10002706 (PIs: D. Kitzmann, H.J. Hoeijmakers) and from the Canadian Space Agency (CSA) under grant 24JWGO3A-03. R.A. acknowledges the Swiss National Science Foundation (SNSF) support under the Post-Doc Mobility grant P500PT\_222212 and the support of the Institut Trottier de Recherche sur les Exoplanètes (IREx). L.-P.C. acknowledges support from Mitacs through the Mitacs Accelerate program, in partnership with the Montreal Planetarium. F.G acknowledges support from the Fonds de recherche du Québec (FRQ) - Secteur Nature et technologies under file $\#$350366. V.Y. acknowledges funding from the physics doctoral school (ED-PHYS) of the Grenoble-Alpes University. Finally, we thank the members of the NIRPS science team for insightful discussions.
\end{acknowledgements}

\bibliographystyle{bibtex/aa} 
\bibliography{citations} 

\begin{appendix}

\color{black}

\section{Parametrization of the line-of-sight velocity field}
\label{app:A_0}

We model the line-of-sight (LOS) velocity field across the planetary disk as the sum of three contributions: solid-body rotation, an equatorial jet, and a source-sink flow. {\color{black}{Let $(x, y, z)$ be Cartesian coordinates in the observer frame, with $\hat{x}$ pointing to the right, $\hat{y}$ pointing upwards, and $\hat{z}$ pointing toward the observer.}} Each point on the visible hemisphere lies on the unit sphere, $x^{2}+y^{2}+z^{2}=1$ with $z>0$, so that $z=\sqrt{1-x^{2}-y^{2}}$ is fully determined by the sky-plane coordinates $(x,y)$. The orbital phase $\phi\in [0,1)$ is measured from mid-transit ($\phi=0$), and the corresponding rotation angle is $2\pi\phi$.

\subsection*{Solid-body rotation}

For a planet in synchronous rotation with equatorial velocity $v_\mathrm{rot}$, the zonal velocity at latitude $\theta$ is $v_\mathrm{rot}\cos\theta$. Since $x=\cos\theta\sin\varphi$ (where $\varphi$ is the longitude), the LOS projection of the rotation velocity is

\begin{equation}
  v_\mathrm{LOS}^{\,\mathrm{rot}}(x,y) = v_\mathrm{rot}\,x.
  \label{eq:v_rot}
\end{equation}

\subsection*{Equatorial jet}

The equatorial jet is a zonal wind with a Gaussian latitudinal profile.  Its
LOS velocity is

\begin{equation}
  v_\mathrm{LOS}^{\,\mathrm{jet}}(x,y) = v_\mathrm{jet}\,x
    \exp\!\left(-\frac{y^{2}}{2\sigma_\mathrm{jet}^{2}}\right),
  \label{eq:v_jet}
\end{equation}

\noindent where $v_\mathrm{jet}$ is the jet speed at zero latitude and $\sigma_\mathrm{jet}$ is the standard deviation of the Gaussian, controlling how rapidly the jet speed decreases away from the equator. The $x$ factor provides the same LOS projection as in Eq. (\ref{eq:v_rot}). The rotation and jet terms share the same morphology but are physically distinct: rotation is a rigid-body term set by the planetary spin, while the jet represents a zonal wind anomaly superimposed on it.

\subsection*{Source-to-sink flow}

The source-to-sink flow has a constant magnitude $v_\mathrm{wind}$ everywhere on the sphere.  It diverges from a source at the substellar point and converges to a sink at longitude $\varphi_\mathrm{sink}$ on the equator, which is a free parameter of the fit.  When $\varphi_\mathrm{sink}=\pm 180^\circ$, the sink coincides with the antistellar point and the flow reduces to a classical day-to-night (i.e., substellar-to-antistellar) wind.

{\color{black}{The source and sink positions are defined in the co-rotating planet frame. However, we need to express them in the observer frame at a given phase $\phi$. By definition, the planet frame coincides with the observer frame at $\phi$ = 0 (mid-transit). With the source fixed at the substellar point ($\varphi_\mathrm{src}=0^\circ$, latitude $0^\circ$, $z_\text{src}= -1$), the planet-frame unit vectors are}

\begin{equation}
  \hat{s}_\mathrm{src} = \begin{pmatrix}0\\0\\-1\end{pmatrix}, \qquad
  \hat{s}_\mathrm{sink} = \begin{pmatrix}-\sin\varphi_\mathrm{sink}\\0\\-\cos\varphi_\mathrm{sink}\end{pmatrix},
  \label{eq:source_sink_planet}
\end{equation}

}

\noindent where longitudes are measured eastward.  These are rotated into the observer frame by the orbital rotation matrix

\begin{equation}
  R(\phi) =
  \begin{pmatrix}
    \cos2\pi\phi & 0 & \sin2\pi\phi \\
    0          & 1 & 0          \\
   -\sin2\pi\phi & 0 & \cos2\pi\phi
  \end{pmatrix},
  \label{eq:rotation_matrix}
\end{equation}

\noindent giving $\hat{s}_\mathrm{src}^{(\mathrm{obs})}=R(\phi)\,\hat{s}_\mathrm{src}$ and $\hat{s}_\mathrm{sink}^{(\mathrm{obs})}=R(\phi)\,\hat{s}_\mathrm{sink}$.

At a surface point $\hat{p}=(x,y,z)$, the angular distances to the source and sink are

\begin{equation}
  \alpha_\mathrm{src}  = \arccos\!\left(\hat{p}\cdot\hat{s}_\mathrm{src}^{(\mathrm{obs})}\right),
  \qquad
  \alpha_\mathrm{sink} = \arccos\!\left(\hat{p}\cdot\hat{s}_\mathrm{sink}^{(\mathrm{obs})}\right).
  \label{eq:angular_distances}
\end{equation}

\noindent The local flow direction is constructed as a weighted combination of the chord vectors pointing away from the source and toward the sink,

\begin{equation}
  \mathbf{f} =
    w_\mathrm{src}\!\left(\hat{p}-\hat{s}_\mathrm{src}^{(\mathrm{obs})}\right)
    + w_\mathrm{sink}\!\left(\hat{s}_\mathrm{sink}^{(\mathrm{obs})}-\hat{p}\right),
  \label{eq:flow_vector}
\end{equation}

\noindent with weights

\begin{equation}
  w_\mathrm{src}  = \frac{\alpha_\mathrm{sink}}{\alpha_\mathrm{src}+\alpha_\mathrm{sink}},
  \qquad
  w_\mathrm{sink} = \frac{\alpha_\mathrm{src}}{\alpha_\mathrm{src}+\alpha_\mathrm{sink}}.
  \label{eq:weights}
\end{equation}

\noindent This weighting ensures that the flow is dominated by the source near the source point and by the sink near the sink point, with equal contributions at the geodesic midpoint. The vector $\mathbf{f}$ is then projected onto the tangent plane of the sphere at $\hat{p}$ by removing its radial component,

\begin{equation}
  \mathbf{t} = \mathbf{f} - \left(\mathbf{f}\cdot\hat{p}\right)\hat{p},
  \label{eq:tangent_projection}
\end{equation}

\noindent and normalized to a unit vector $\hat{t}=\mathbf{t}/|\mathbf{t}|$, ensuring that the wind is everywhere tangent to the planetary surface and has a constant magnitude $v_\mathrm{wind}$. The LOS velocity contribution is then

\begin{equation}
  v_\mathrm{LOS}^{\,\mathrm{wind}}(x,y;\phi) = -v_\mathrm{wind}\,\hat{t}\cdot\hat{z}
    = -v_\mathrm{wind}\,t_{z},
  \label{eq:v_wind}
\end{equation}

\noindent where the negative sign follows the convention that a flow component toward the observer ($t_{z}>0$) produces a blueshift.

\subsection*{Total velocity field}

In summary, the total LOS velocity field at sky-plane position $(x,y)$ and orbital phase
$\phi$ is

\begin{equation}
  v_\mathrm{LOS}(x,y;\phi) =
    v_\mathrm{rot}\,x
    + v_\mathrm{jet}\,x\exp\!\left(-\frac{y^{2}}{2\sigma_\mathrm{jet}^{2}}\right)
    + v_\mathrm{LOS}^{\,\mathrm{wind}}(x,y;\phi).
  \label{eq:v_total}
\end{equation}

\noindent The velocity field has five parameters in total: the equatorial rotation velocity
$v_\mathrm{rot}$, the jet speed $v_\mathrm{jet}$ and width $\sigma_\mathrm{jet}$,
the wind speed $v_\mathrm{wind}$, and the sink longitude
$\varphi_\mathrm{sink}$. The phase dependence of the velocity field enters exclusively through the
source-to-sink flow, via the rotation of $\hat{s}_\mathrm{src}^{(\mathrm{obs})}$
and $\hat{s}_\mathrm{sink}^{(\mathrm{obs})}$ under angle $2\pi\phi$.

\section{Parametrization of the more complex weight function}
\label{app:A_1}

\begin{figure*}[t!]
        \centering
        \vspace{0pt}
        \makebox[\textwidth][c]{\hspace{-10pt}\includegraphics[width=0.83\textwidth]{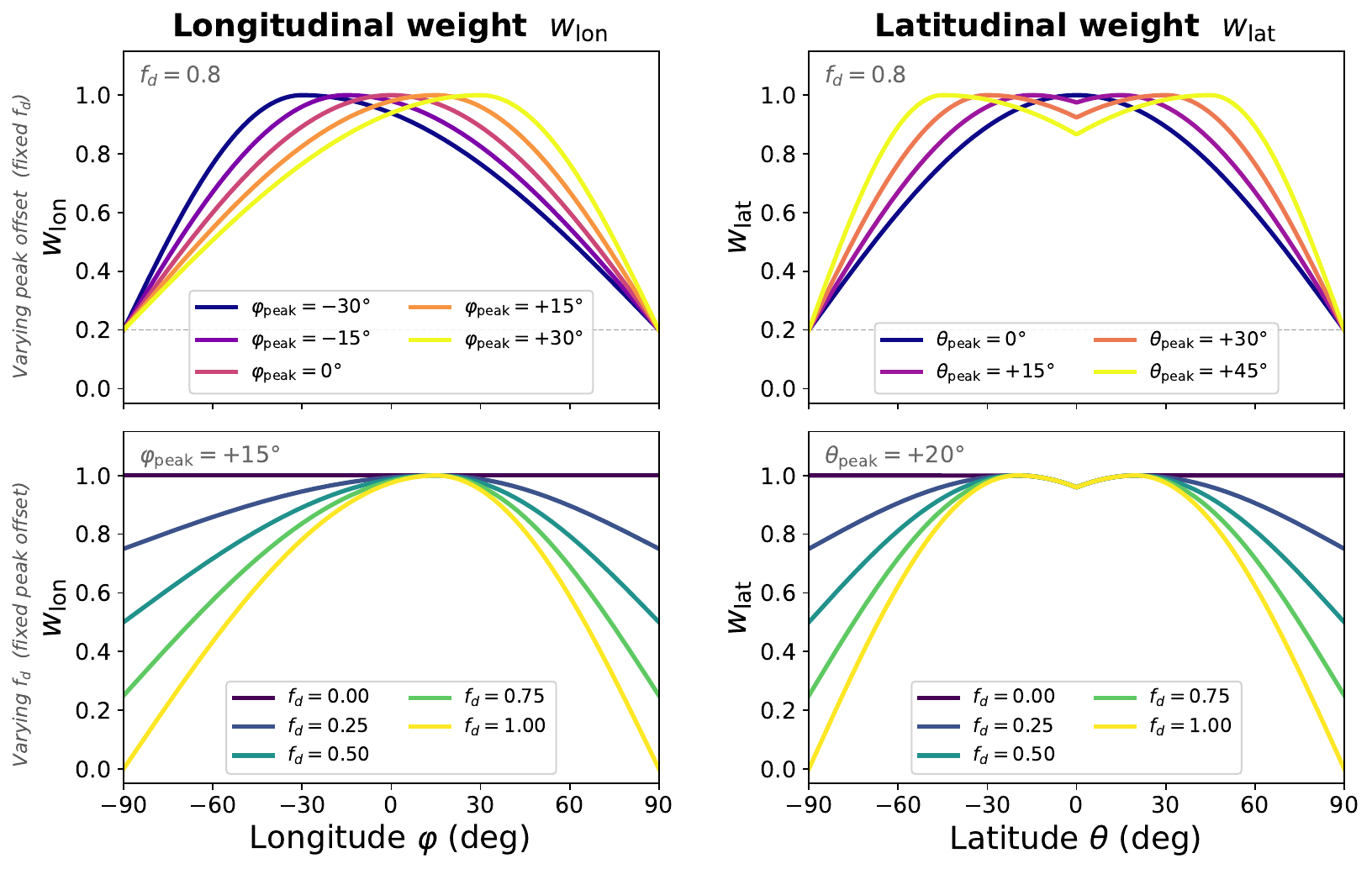}}
        \vspace{-10pt}
        \caption{{\color{black}{Visualization of the longitude dependence (left column) and the latitude dependence (right column) of the more complex weight function described in Applendix \ref{app:A_1}. In the top row, we vary the latitudinal and longitudinal peak offsets while keeping the dropoff parameter $f_\text{d}$ fixed at 0.8. In the bottom row, we vary the value of the dropoff parameter, while keeping the longitudinal/latitudinal peak offset fixed at 15/20 degrees.}}}
        \label{fig:weight_mask_visualisation}
\end{figure*}

In high-resolution emission spectroscopy, the observed line-broadening profile is a weighted histogram of the LOS velocities across the planetary disk, where each surface element contributes with a weight proportional to the local line strength.  The line strength is set by the flux difference between the line core and the continuum, rather than by the absolute flux, and the weight function should be interpreted accordingly. For this reason, we do not use terms like ``limb darkening'' or ``hotspot'', as these tend to be associated with absolute flux.

We parametrize the total weight function $w_\mathrm{tot}$ as a product of three independent contributions:

\begin{equation}
  w_\mathrm{tot}(x, y;\, \phi) = w_\mathrm{lon}(\varphi)\cdot w_\mathrm{lat}(\theta)\cdot w_\mathrm{rad}(\mu),
  \label{eq:weight_total}
\end{equation}

\noindent where $\phi$ is the orbital phase, $\varphi$ denotes the longitude, $\theta$ is the latitude, and $\mu = z = \sqrt{1-x^2-y^2}$ is the cosine of the angle between the surface normal and the line of sight. Note that $\mu$ is defined in the observer frame, while $\varphi$ and $\theta$ are coordinates fixed on the planetary sphere. By definition, we assume that the weight function is zero on the nightside, for which $|\varphi|>\pi/2$. Also, since only the \emph{relative} weights matter for the construction of the Doppler kernel, the overall normalization of $w_\mathrm{tot}$ is arbitrary. We describe the three components from Eq. (\ref{eq:weight_total}) in the sections below. 

\subsection*{Longitude dependence}

The longitudinal weight is a piecewise cosine centered on a longitude $\varphi_\mathrm{peak}$. As such, $\varphi_\mathrm{peak} > 0$ implies an eastward ``weight shift'', while $\varphi_\mathrm{peak} < 0$ implies a westward ``weight shift''. The \emph{observer-frame} longitude associated with a point $\hat{p}=(x,y,z)$ on the planetary sphere is

\begin{align}
  \varphi_\mathrm{obs} &= \arctan\!\left(\frac{x}{z}\right) \in \left[-\tfrac{\pi}{2}, \tfrac{\pi}{2}\right].
\end{align}

\noindent For a given orbital phase $\phi$, the planet-frame longitude $\varphi$ that corresponds to $\varphi_\mathrm{obs}$ is 

\begin{equation}
  \varphi = \left(\varphi_\mathrm{obs} - 2\pi\phi\right) \bmod 2\pi - \pi.
  \label{eq:corot_lon}
\end{equation}

\noindent Defining $\varphi_\mathrm{rel} \equiv \varphi - \varphi_\mathrm{peak}$, the longitudinal
weight is

\begin{equation}
  w_\mathrm{lon}(\varphi) =
  \begin{cases}
    f_\mathrm{d}\cos\!\left(\dfrac{\pi}{2}\,\dfrac{\varphi_\mathrm{rel}}{L_\mathrm{w}}\right) + (1 - f_\mathrm{d})
      & \varphi_\mathrm{rel} \leq 0\ \text{(west of peak)}, \\[10pt]
    f_\mathrm{d}\cos\!\left(\dfrac{\pi}{2}\,\dfrac{\varphi_\mathrm{rel}}{L_\mathrm{e}}\right) + (1 - f_\mathrm{d})
      & \varphi_\mathrm{rel} > 0\ \ \text{(east of peak)}, 
  \end{cases}
  \label{eq:lon_weight}
\end{equation}

\noindent where $L_\mathrm{w} = \varphi_\mathrm{peak} + \pi/2$ and $L_\mathrm{e} = \pi/2 - \varphi_\mathrm{peak}$ are the angular distances from the peak to the western and eastern terminators, respectively. The dropoff parameter $f_\mathrm{d}\in[0,1]$ controls the contrast between the peak and the terminators: when $f_\mathrm{d}=0$ the dayside is uniformly weighted in longitude, and when $f_\mathrm{d}=1$ the weight drops to zero at both terminators. Furthermore, the weight is unity at the peak by construction. The left panels in Fig. \ref{fig:weight_mask_visualisation} show what the longitude dependence of the weight function looks like for different values of the peak offset and the dropoff parameter.

\subsection*{Latitude dependence}

The latitudinal weight has two symmetric peaks at latitudes $\pm\theta_\mathrm{peak}$ (with $\theta_\mathrm{peak}\geq 0$), each following a cosine roll-off toward the nearest pole. The combined weight is the maximum of the contributions from the two peaks,

\begin{equation}
  w_\mathrm{lat}(\theta) = \max\!\left[w_+(\theta),\, w_-(\theta)\right],
  \label{eq:lat_weight_combined}
\end{equation}

\noindent where $w_+$ and $w_-$ correspond to the northern and southern peaks, respectively. Also, $w_-(\theta)=w_+(-\theta)$ by symmetry. For the northern peak, defining $\theta_\text{rel} \equiv \theta - \theta_\text{peak}$, we have

\begin{equation}
  w_+(\theta) =
  \begin{cases}
    \cos\!\left(\dfrac{\pi}{2}\,\dfrac{\theta_\text{rel}}{L_\mathrm{s}}\right)
      & \theta_\text{rel} \leq 0\ \ \text{(south of peak)}, \\[10pt]
    f_\mathrm{d}\cos\!\left(\dfrac{\pi}{2}\,\dfrac{\theta_\text{rel}}{L_\mathrm{n}}\right)
      + (1 - f_\mathrm{d})
      & \theta_\text{rel} > 0\ \ \text{(north of peak)},
  \end{cases}
  \label{eq:lat_weight_north}
\end{equation}

\noindent with $L_\mathrm{n} = \pi/2 - \theta_\mathrm{peak}$ (peak to north pole) and $L_\mathrm{s} = \pi/2 + \theta_\mathrm{peak}$ (peak to south pole). The poleward roll-off uses the same dropoff floor $1-f_\mathrm{d}$ as the longitudinal weight (to minimize the number of parameters), while the equatorward roll-off always reaches zero at the opposite pole, ensuring the two peaks remain distinct.

\subsection*{Radial dependence}

The radial weight accounts for center-to-limb variations on the disk that are not dependent on latitude and longitude. This is the weight function we also use in the nominal retrievals. We adopt a linear law in $\mu$,

\begin{equation}
  w_\mathrm{rad}(\mu) = 1 + (w_0 - 1)\,\mu,
  \label{eq:rad_weight}
\end{equation}

\noindent where $w_0$ is the weight at the disk center relative to the limb. By construction, the weight at the limb ($\mu = 0$) is always unity, while $w_0$ sets the relative weight at disk center. As such, $w_0>1$ corresponds to a more ``center-weighted'' profile, while $w_0<1$ produces a more ``limb-weighted'' profile. We adopt this parametrization -- fixing the limb weight to unity rather than the disk center -- because the retrievals performed in this work preferred more `limb-weighted'' broadening kernels. If we set the weight of the center to unity instead (as is done in a classical limb-darkening parametrization), the weight at the limb could potentially become unbounded. 

\subsection*{Doppler broadening kernel}

The Doppler broadening kernel $\vec{k}$ is constructed as the weight-integrated histogram of LOS velocities across the visible disk,

\begin{equation}
  k(v;\phi) \propto \iint_{x^2+y^2\leq 1}
    w\!\left(x, y;\phi\right)\,
    \delta\!\left[v - v_\mathrm{LOS}(x,y;\phi)\right]
    \,\mathrm{d}x\,\mathrm{d}y,
  \label{eq:kernel}
\end{equation}

\noindent where $v_\mathrm{LOS}$ is the velocity field defined in Appendix~\ref{app:A_0} and $\delta$ is the Dirac delta function. In practice, the integral is evaluated as a discrete sum over the grid points of the planetary disk.  Our complex weight function has four parameters in total: the longitudinal peak shift $\varphi_\mathrm{peak}$, the latitudinal peak shift $\theta_\mathrm{peak}$, the dropoff parameter $f_\mathrm{d}$, and the center-to-limb weight $w_0$.

\section{Supplementary figures}
\label{app:B}

\begin{figure*}
\begin{minipage}{\textwidth}
    \centering
    \vspace{-13pt}
    \makebox[\textwidth][c]{\includegraphics[width=0.9\textwidth]{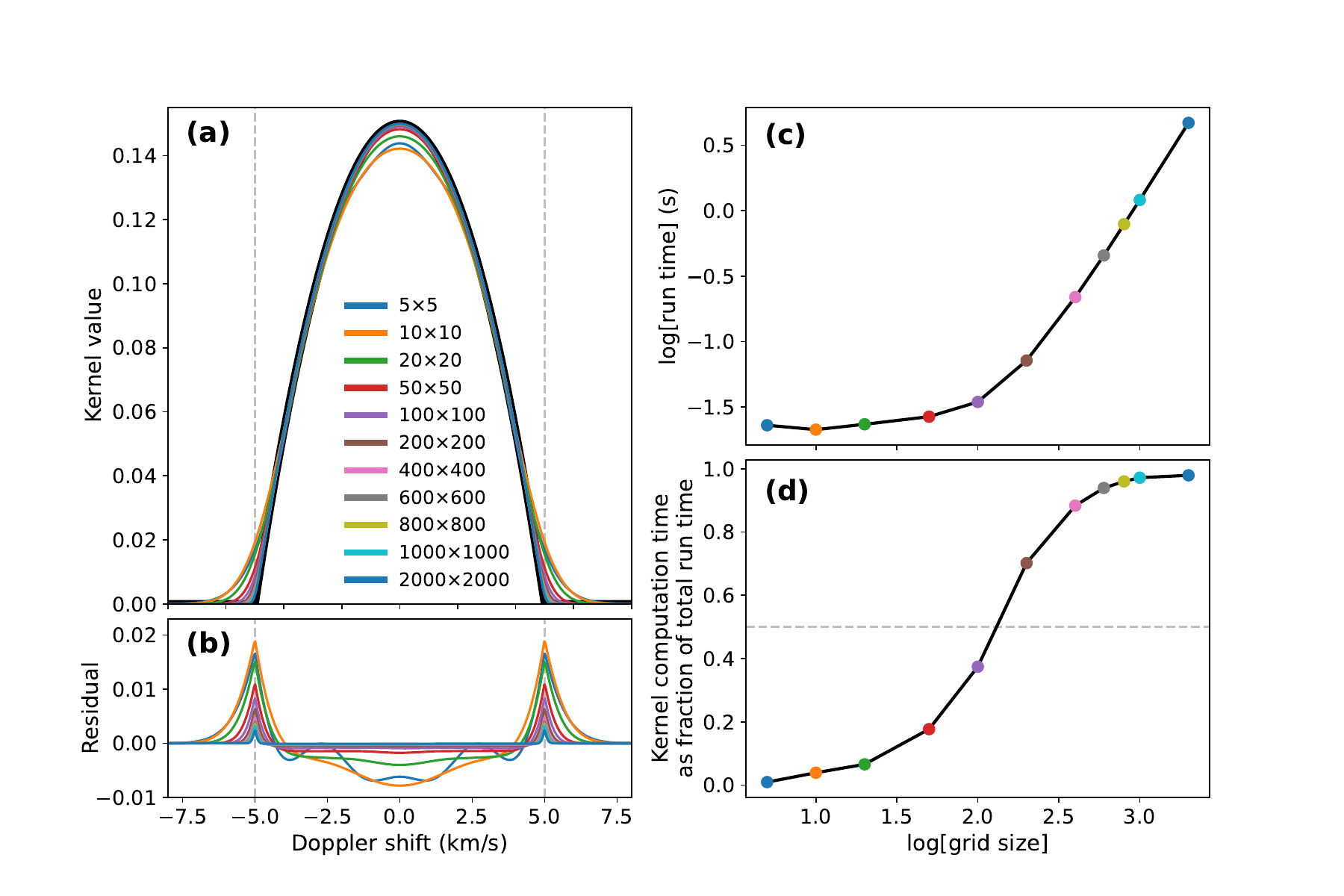}}
    \vspace{-28pt}
    \caption{Tests demonstrating the performance of \texttt{dopplerkernel}, the forward model used in our retrievals. Panel (a): Broadening kernels for solid body rotation with $v_\text{rot} = $ 5 km/s and a linear limb-darkening coefficient $u_1$ = 1 (assuming the traditional limb-darkening parametrization). The colored curves show the kernels obtained for different grid sizes (denoted as $n_\text{cells}\times n_\text{cells}$). The thicker back curve in the background shows the analytical solution (e.g., \citealt{Gray2005}). Panel (b): Residuals obtained by subtracting the analytical solution from the numerical broadening kernels. \mbox{Panel (c):} Run time of a single forward model as a function of the grid size (denoted as log[$n_\text{cells}$]) for an input spectrum with 100,000 wavelength points. Running a forward model involves: calculating the velocity field, calculating the weight function, computing the broadening kernel, and convolving the input spectrum with the broadening kernel (see Section \ref{sec:methods}). Panel (d): Fraction of the run time spent on computing the broadening kernel as a function of grid size. This test was run on a single core of an Apple M4 Pro chip.}
    \label{fig:app_A}
\end{minipage}

    \vspace{20pt}

    \begin{minipage}[t]{0.495\textwidth}
        \centering
        \includegraphics[width=\textwidth]{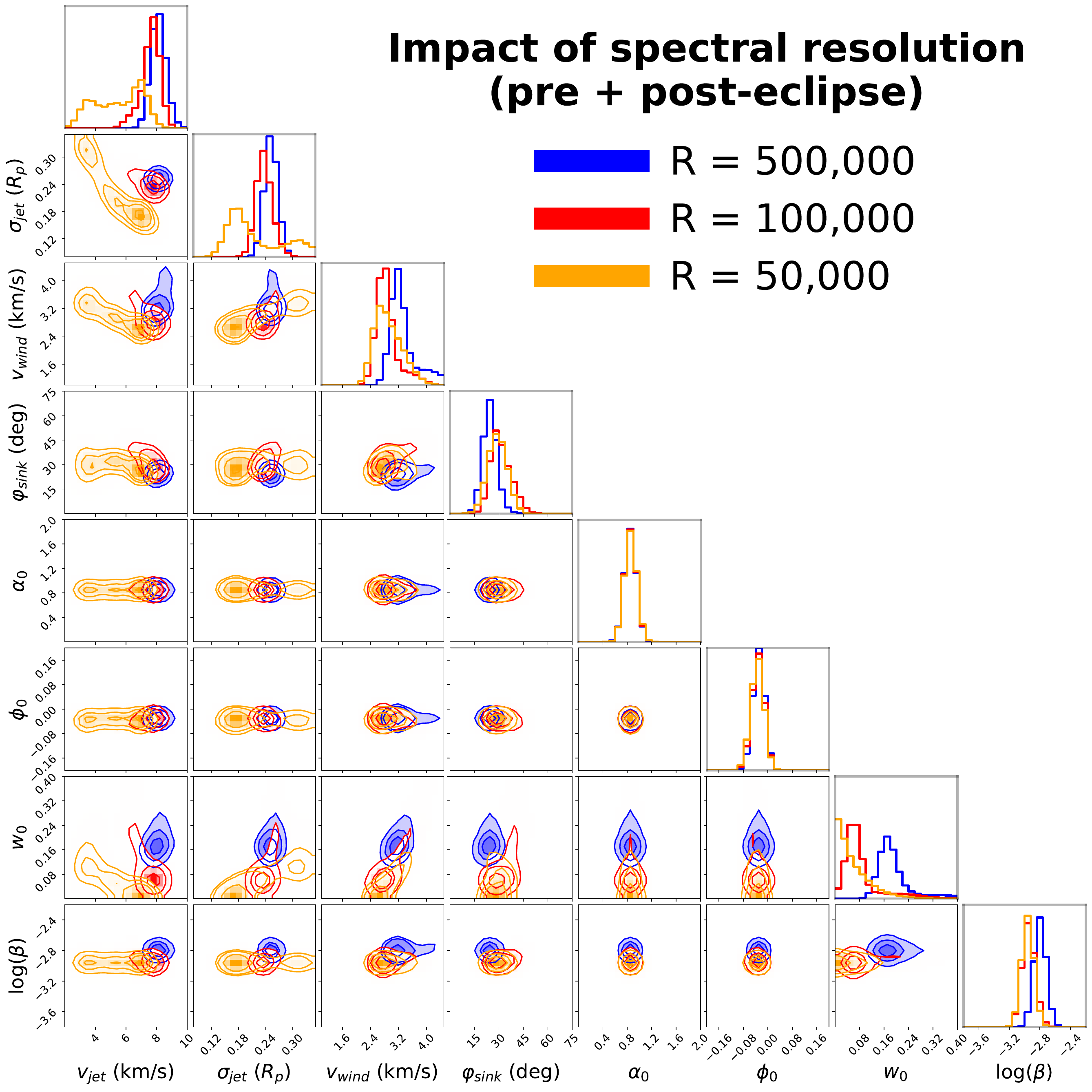}
    \end{minipage}
    \hfill
    \begin{minipage}[t]{0.495\textwidth}
        \centering
        \includegraphics[width=\textwidth]{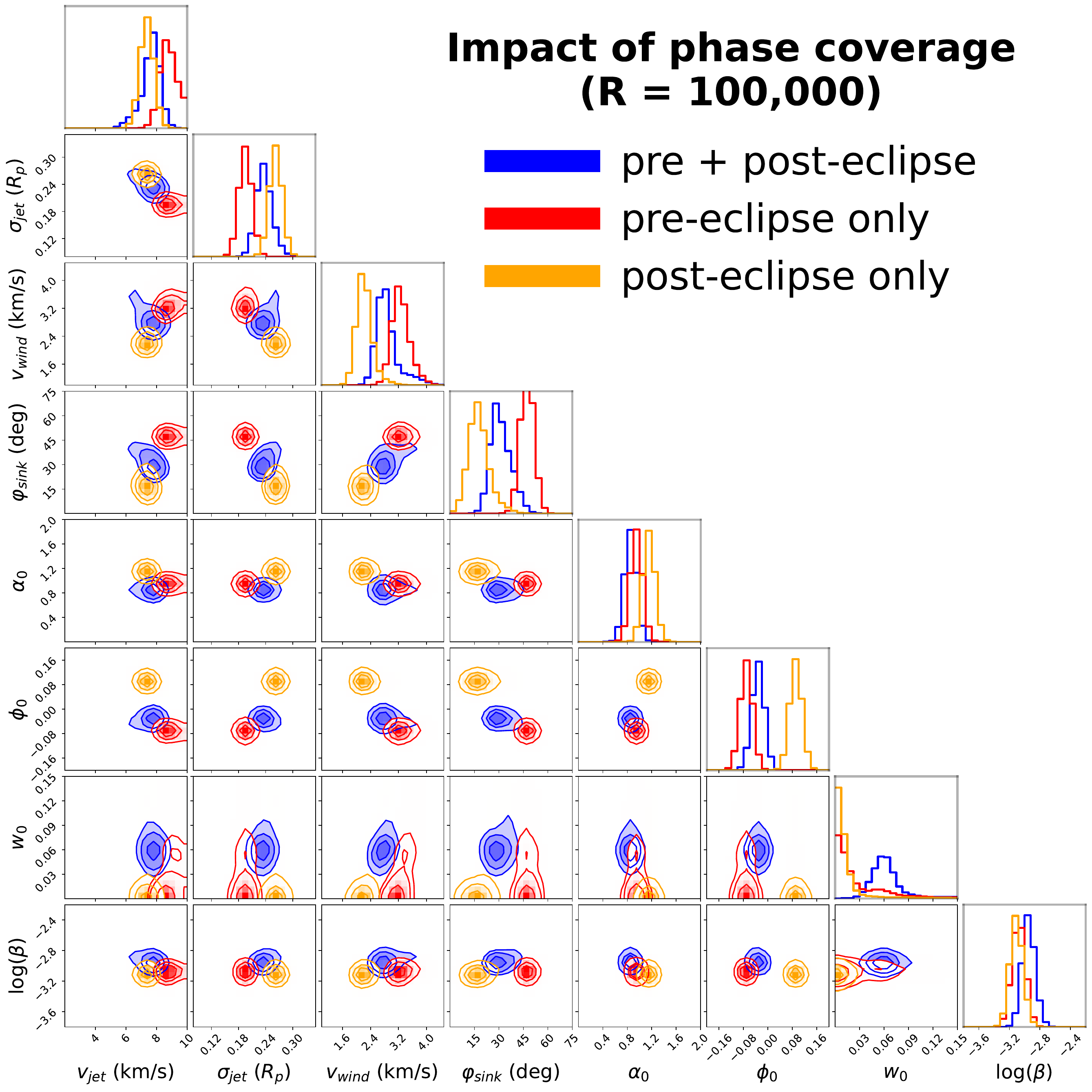}
    \end{minipage}
    \vspace{-3pt}
    \caption{{\color{black}{Same corner plots as in Fig \ref{fig:two_panel_fullwidth}, but now showing the posteriors for all the parameters included in the retrieval.}}}
    \label{fig:appendix_full_corner_pair}
\end{figure*}

Fig. \ref{fig:app_A} shows the accuracy of the Doppler-broadening model as a function of grid size for the case of solid-body rotation with linear limb darkening. Additionally, we plot the total run time of the forward model, as well as the time spent calculating the kernel as a fraction of the total model run time.

Fig. \ref{fig:appendix_full_corner_pair} shows the full corner plots for the retrievals discussed in Section \ref{ref:impact_reso_phases}, in which we investigated the impact of spectral resolution and phase coverage on the wind measurements (a small version of the same plot is shown in Fig. \ref{fig:two_panel_fullwidth}).  

\begin{figure*}
        \centering
        \makebox[\textwidth][c]{\includegraphics[width=0.7\textwidth]{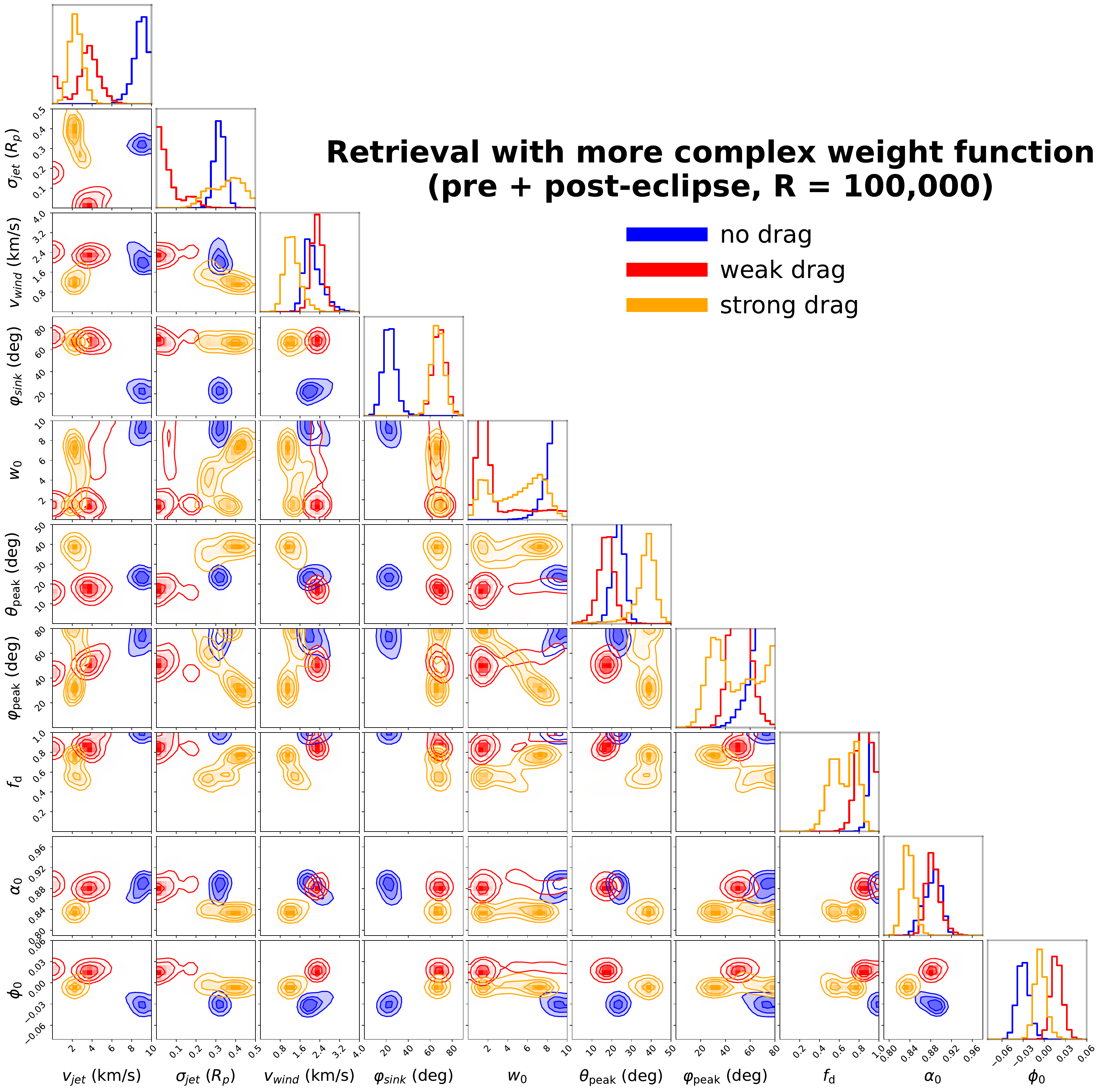}}
        \vspace{-12pt}
        \caption{Full corner plot for the retrieval with the more complex weight function discussed in Section \ref{sec:advanced_weight_function}.}
        \label{fig:app_B}

        \vspace{-10pt}

        \begin{minipage}{\textwidth}
            \centering
            \makebox[\textwidth][c]{\hspace{-10pt}\includegraphics[width=0.85\textwidth]{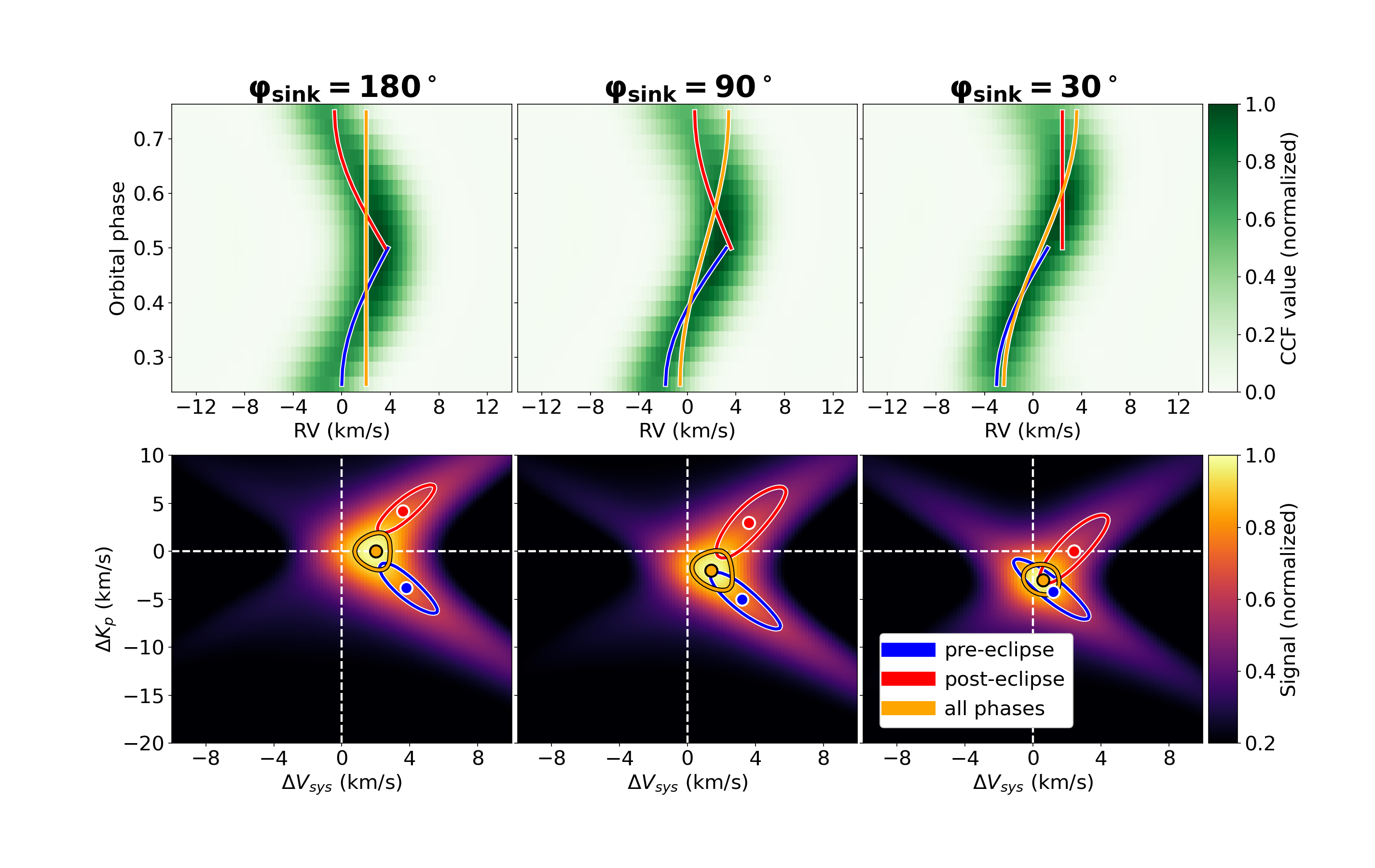}}
            \vspace{-34pt}
            \caption{Same type of plot as Fig. \ref{fig:ccf_and_kp_vsys_maps}, but now we examine the impact the sink longitude $\varphi_\mathrm{sink}$ on the $K_\text{p}$ and $V_\text{sys}$ offsets induced by the source-to-sink flow. In each model, we assume $v_\text{rot} = v_\text{jet} = 0$ km/s, $v_\text{wind}$ = 4 km/s, $w_0$ = 0.2, $\alpha_0$ = 1, and $\phi_0 = $ 0. The only parameter that is changing between the three columns is $\varphi_\mathrm{sink}$. In the case of a symmetric source-to-sink flow (with $\varphi_\text{sink} = \pm 180^\circ$), the CCF trail is symmetric about phase 0.5, and the best-fit-trail across all phases is a straight line with a positive offset (implying $\Delta K_\text{p} = 0$ and $\Delta V_\text{sys} > 0$). However, shifting the sink away from the antistellar point causes the CCF trail to become more asymmetric about phase 0.5, such that trail with a nonzero $\Delta K_\text{p}$ is needed to fit the phase-dependent Doppler shifts.}
            \label{fig:app_C}
        \end{minipage}
\end{figure*}

Fig. \ref{fig:app_B} shows the full corner plot for retrieval with the more complex weight function discussed in Section \ref{sec:advanced_weight_function} (a small version of the same plot is shown in Fig. \ref{fig:corner_hotspot_offset}). 

Fig. \ref{fig:app_C} illustrates the impact of the sink longitude $\varphi_\mathrm{sink}$ on the $K_\text{p}$ and $V_\text{sys}$ offsets induced by the source-to-sink flow.

\end{appendix}
\end{document}